\documentclass[%
 aip,
 amsmath,amssymb,
 reprint,%
]{revtex4-1}

\usepackage{graphicx}
\usepackage{bm}
\usepackage{dcolumn}
\usepackage{hyperref}
\begin{document}

\preprint{AIP/123-QED}

\title{Spectral Analysis of Brownian Motion with its Rheological Analogues}

\author{Nicos Makris}
\affiliation{Dept. of Civil and Environmental Engineering, Southern Methodist University, Dallas, Texas, 75276}

\date{\today}

\begin{abstract}
The power spectrum of the Brownian motion of probe microparticles with mass $m$ and radius $R$ immersed in a viscoelastic material reveals valuable information about repetitive patterns and correlation structures that manifest in the frequency domain. In this paper, we employ a viscous--viscoelastic correspondence principle for Brownian motion and we show that the power spectrum (power spectral density) of Brownian motion in any linear, isotropic viscoelastic material is proportional to the real part of the complex dynamic fluidity (complex mobility) of a linear rheological network that is a parallel connection of the linear viscoelastic material within which the Brownian particles are immersed and an inerter, with distributed intrance with mass $m_R = \frac{m}{6\pi R}$. The synthesis of this rheological analogue simplifies appreciably the calculation of the power spectrum for Brownian motion within viscoelastic materials such as Maxwell fluids, Jeffreys fluids, subdiffusive materials, or in dense viscous fluids that give rise to hydrodynamic memory.
\end{abstract}

\maketitle

\section*{Introduction}

Thermally driven Brownian motion of probe microparticles immersed within a viscoelastic material is the result of the perpetual random collisions of the molecules of the surrounding material to the probe microparticles (more than $10^{15}$ collisions per second for a $1\,\mu\mathrm{m}$ $(10^{-6} m)$ diameter microparticle). This large number of collisions induce random fluctuations of the position and velocity of the Brownian particles and their position, $x$ at any given time $t$ is described by a probability density function, $p(x,t)$ as was first shown by Einstein~\cite{einstein1905}, upon solving a one-dimensional diffusion equation which predicted the long term (diffusing regime) of Brownian particles immersed in a memoryless, Newtonian viscous fluid. Einstein's~[1] long term solution shows that the variance, $\sigma^2(t)$ of the position probability density $p(x,t)$ grows linearly with time, $\sigma^2(t)=2Dt$, where $D$ is a time-independent diffusion coefficient of the Brownian process within the memoryless material. Given that the distribution, $p(x,t)$ of the Brownian process spreads linearly with time ($\sigma^2(t)=2Dt$), Brownian motion in a memoryless, Newtonian viscous fluid is not a stationary random process. Nevertheless, it is a process with stationary increments which implies that the evolution of the Gaussian distribution, $p(x,t)$ and correlations during the Brownian process over a given time interval are the same, regardless when the time interval starts. As an example for an ensemble (collection) of $M$ Brownian particles and two distinct times $\xi_1$ and $\xi_2$ ($\xi_1 \neq \xi_2$) the ensemble average velocity correlation function
\begin{equation}
\begin{aligned}
\langle v(\xi_1)v&(\xi_1+t)\rangle = \frac{1}{M}\sum_{j=1}^{M} v_j(\xi_1)\,v_j(\xi_1+t) \\
&= \frac{1}{M}\sum_{j=1}^{M} v_j(\xi_2)\,v_j(\xi_2+t)
= \langle v(\xi_2)v(\xi_2+t)\rangle ,
\end{aligned}
\label{eq:velcorr}
\end{equation}
is independent on whether one starts at time $\xi_1$ or time $\xi_2$.

Similarly, the position probability density function $p(x,t)$ of Brownian particles trapped in a harmonic potential well with dissipation (damped harmonic oscillator) can be calculated by solving the more elaborate Fokker--Planck equation~\cite{pathria1996,risken1996,araujo2012,polotto2018,santra2001},
, which also leads to a Gaussian position probability density function which initially spreads with time and eventually reaches a final shape with a time independent variance, 
$\sigma^2_{t=large} = \dfrac{k_B T}{6 \pi R G}$~\cite{pathria1996}, 
where $k_B$ is Boltzmann’s constant, $T$ is the equilibrium temperature of the material surrounding the Brownian microspheres with radius $R$, and $G$ is the elastic shear modulus of the surrounding solid-like material. Accordingly, given that the underlying mechanism (process) which generates Brownian motion from the collisions of molecules is stationary (does not change on an average with time) in association that the Brownian motion outcome has stationary increments as expressed by Eq.~(1); one can proceed with spectral analysis of Brownian motion as was brought forward in the seminal 1945 paper by Wang and Uhlenbeck~\cite{wang1945}.

Soon after Einstein’s 1905 paper~\cite{einstein1905} in which Brownian motion was explained by solving a one-dimensional diffusion equation that produced the Gaussian position probability density function of the Brownian particles; Langevin~\cite{langevin1908} explained Brownian motion by adopting an entirely different approach~\cite{landau1980,attard2012,coffey2012}. In his 1908 paper, Langevin mingles statistical and continuum mechanics to formulate the equation of motion of a Brownian microsphere with radius $R$ and mass $m$, suspended in a Newtonian viscous fluid with shear viscosity $\eta$ when subjected to the random forces $f_R(t)$ that originate from the collision of the fluid molecules on the Brownian microsphere:
\begin{equation}
m\,\frac{dv(t)}{dt} = -\zeta\,v(t) + f_R(t),
\label{eq:langevin1}
\end{equation}
where $v(t) = \frac{dr(t)}{dt}$ is the particle velocity and $\zeta v(t)$ is a viscous drag force proportional to the velocity of the Brownian particle. For a memoryless, viscous fluid with shear viscosity $\eta$, the drag coefficient is given by Stokes law $\zeta = 6 \pi R \eta$\cite{landau1980fluid}.

Upon dividing with the mass $m$ of the Brownian microsphere, Eq.~(\ref{eq:langevin1}) assumes the expression
\begin{equation}
\frac{dv(t)}{dt} + \frac{1}{\tau}v(t) = \frac{f_R(t)}{m},
\label{eq:langevin2}
\end{equation}
where $\tau = \frac{m}{6\pi R \eta}$ is the dissipation time of the perpetual fluctuation–dissipation process. The random excitation $f_R(t)$ has a zero average value over time, $\langle f_R(t) \rangle = 0$, while for the memoryless viscous fluid that only dissipates energy (no elasticity), the force correlation function contracts to a Dirac delta function~\cite{lighthill1958}

\begin{equation}
\langle f_R(t_1) f_R(t_2) \rangle = A\,\delta(t_1 - t_2)
\tag{4}\label{eq:forcecorr}
\end{equation}
with $t_1 \neq t_2$ and $A$, a constant that expresses the strength of the random forces.

Given the random nature of the excitation force $f_R(t)$, the Langevin Eq.~(3) can be integrated
in terms of ensemble averages in association with Eq.~(\ref{eq:forcecorr})~\cite{landau1980,attard2012,coffey2012} and the mean-square displacement of microparticles suspended in a viscous fluid was first computed by Ornstein~\cite{ornstein1917}:
\begin{equation}
\begin{aligned}
\left\langle \Delta r^{2}(t) \right\rangle
&= \frac{1}{M}\sum_{j=1}^{M} \bigl(r_j(t)-r_j(0)\bigr)^2 \\
&= \frac{N k_B T}{3\pi R}\,\frac{1}{\eta}
\Bigl[t - \tau\bigl(1-e^{-t/\tau}\bigr)\Bigr] .
\end{aligned}
\tag{5}
\end{equation}
where $N\in\{1,2,3\}$ is the number of spatial dimensions, while $r_j(t)$ and $r_j(0)$ are the positions of particle $j$ at time $t$ and at the time origin, $t=0$.

The Laplace transform of the mean-square displacement, $\mathcal{L}\{\langle \Delta r^{2}(t) \rangle\}
= \langle \Delta r^{2}(s) \rangle
= \int_{0}^{\infty} \langle \Delta r^{2}(t) \rangle e^{-st} \, dt,$ is related to the Laplace transform of the velocity autocorrelation function $\mathcal{L}\{\langle v(0)v(t) \rangle\}
= \langle v(0)v(s) \rangle
= \int_{0}^{\infty} \langle v(0)v(t) \rangle e^{-st} \, dt,$ 
via the identity~\cite{attard2012,coffey2012,squires2010}
\begin{equation}
\langle v(0)v(s) \rangle = \frac{s^{2}}{2}\,\langle \Delta r^{2}(s) \rangle
\tag{6}\label{eq:laplace_identity}
\end{equation}
while, according to the properties of the Laplace transform of the derivatives of a function,
\begin{equation}
s^{2}\langle \Delta r^{2}(s) \rangle
= \mathcal{L}\left\{
\frac{d^{2}\langle \Delta r^{2}(t)\rangle}{dt^{2}}
\right\}
 +s\,\langle \Delta r^{2}(0)\rangle
+ \frac{d\langle \Delta r^{2}(0)\rangle}{dt}.
\tag{7}\label{eq:laplace_derivative}
\end{equation}
From Eq.~(5), at the time origin $t=0$, $\langle\Delta r^{2}(0)\rangle=0$. Furthermore, the time-derivative of the left-hand side of Eq.~(5), which holds for Brownian motion of microparticles suspended in any material gives  
\begin{equation}
\frac{d\langle \Delta r^{2}(t) \rangle}{dt}
= \frac{2}{M}\sum_{j=1}^{M} \big(r_j(t) - r_j(0)\big)\frac{dr_j(t)}{dt}.
\tag{8}
\label{eq:dmsd_dt}
\end{equation}

Consequently, at $t = 0$, from equation~(8), $\dfrac{d\langle \Delta r^2(t) \rangle}{dt} = 0,$ and substitution of Eq.~(\ref{eq:laplace_derivative}) into Eq.~(\ref{eq:laplace_identity}) gives
\begin{equation}
\mathcal{L}\{\langle v(0)v(t)\rangle\}
= \frac{1}{2}\,\mathcal{L}\left\{
\frac{d^{2}\langle \Delta r^{2}(t)\rangle}{dt^{2}}
\right\}.
\tag{9}
\label{eq:laplace_relation}
\end{equation}
The inverse Laplace transform of Eq.~(\ref{eq:laplace_relation}) yields
\begin{equation}
\langle v(0)v(t)\rangle
= \frac{1}{2}\,\frac{d^{2}\langle \Delta r^{2}(t)\rangle}{dt^{2}},
\tag{10}
\label{eq:vacf_msd}
\end{equation}
which shows that the velocity autocorrelation function is half the second time-derivative of the mean-square displacement~\cite{kenkre1981,bian2016}.
The time derivative of the right-hand side of Eq.~(5) is
\begin{equation}
\frac{d\langle \Delta r^{2}(t)\rangle}{dt}
= \frac{N\,k_{B}T}{3\pi R}\, \frac{1}{\eta}
\big(1 - e^{-t/\tau}\big),
\tag{11}\label{eq:dmsd_dt_full}
\end{equation}
indicating that at $t=0$, $\dfrac{d\langle \Delta r^{2}(t)\rangle}{dt} = 0$, which is in agreement with the result of Eq.~(\ref{eq:dmsd_dt}).  
Equation~(10), in association with the result of Eq.~(11), 
yields the velocity autocorrelation function of Brownian particles with mass $m$ when suspended in a memoryless, Newtonian fluid with viscosity~$\eta$:
\begin{equation}
\langle v(0)v(t)\rangle
= \frac{1}{2}\,\frac{d^{2}\langle \Delta r^{2}(t)\rangle}{dt^{2}}
= \frac{N k_{B}T}{m}\,e^{-t/\tau}.
\tag{12}\label{eq:vacf_exp}
\end{equation}
which is the classical result derived by~\cite{wang1945,uhlenbeck1930} after evaluating ensemble averages of the random Brownian process.  
Equation~(\ref{eq:vacf_exp}), while valid for all time scales, does not account for the hydrodynamic memory that manifests as the energized Brownian particle displaces a dense fluid in its immediate vicinity~\cite{zwanzig1970,widom1971,hinch1975,clercx1992,franosch2011,jannasch2011,makris2021}.

The reader recognizes that the exponential term of the velocity autocorrelation function 
given by Eq.~(\ref{eq:vacf_exp}) is whatever is left after taking the second time derivative 
of the mean-square displacement given by Eq.~(5) that is valid for all time scales. 
Consequently, by accounting for the "ballistic regime" at short time scales, Ornstein's 1917~\cite{ornstein1917} expression for the mean-square displacement given by Eq.~(5), is consistent with the identity given by Eq.~(\ref{eq:vacf_msd}), and indicates that the velocities of Brownian particles 
suspended in a memoryless Newtonian fluid are correlated only because of the ballistic regime. In contrast, Einstein’s 1905 "long-term" expression for the mean-square displacement, $\langle r^{2}(t) \rangle = 2N D t,$ 
(diffusive regime) yields an invariably zero velocity autocorrelation function.

\section*{Power Spectrum and Correlations}

Similar to the velocity autocorrelation function given by Eq.~(1), 
and for the specific case where the Brownian particles are immersed in a memoryless, viscous fluid that
is given by Eq.~(12), the power spectrum of a time-domain process (signal) reveals the pressence of repetitive patterns 
and correlation structures of the process in the frequency domain. 
The power spectrum (power spectral density) of a signal is the squared modulus of its Fourier transform~\cite{wang1945,Papoulis1962,Bracewell1965}, therefore it is a real-valued function in the frequency domain, 
and is the Fourier transform of the two-sided autocorrelation function (even function) of the signal.

The ensemble-average velocity autocorrelation function given by Eq.~(\ref{eq:vacf_exp}) 
is a one-sided time-domain function starting at $t=0$, therefore, its Fourier transform is a complex-valued quantity:
\begin{equation}
\begin{aligned}
&\text{VAC}(\omega)
= \frac{1}{2\pi} \int_{-\infty}^{\infty}
\langle v(0)v(t) \rangle e^{-i\omega t}\, dt \\
&= \frac{N k_B T}{m} \frac{1}{2\pi}
\int_{0}^{\infty} e^{-t/\tau} e^{-i\omega t}\, dt = \frac{N k_B T}{6 \pi R \eta}\,
\frac{\tau}{1 + i\omega \tau}
\end{aligned}
\tag{13}\label{eq:power_spectrum}
\end{equation}
where $\tau = \dfrac{m}{6\pi R \eta}$ is the dissipation time of the fluctuation–dissipation process. 
Nevertheless, given that the Brownian motion process has stationary increments, 
the ensemble-average velocity correlations $\langle v(\xi)v(\xi+t) \rangle$ 
during a time interval that starts at time $\xi>0$, shall be the same with the ensemble average velocity correlations
$\langle v(\xi)\,v(\xi-t)\rangle$ that happened before the initiation of the
time interval at time $\xi$. Accordingly, the one–sided velocity
autocorrelation function given by Eq.~(12) can be expressed as a
two–sided even function:
\begin{equation}
\langle v(0)v|t| \rangle = \frac{N\,k_{B}T}{m}\,e^{-|t|/\tau}.
\tag{14}\label{eq:two_sided_corr}
\end{equation}

The Fourier transform of the two-sided even (symmetric) autocorrelation function 
given by Eq.~(\ref{eq:two_sided_corr}) is~\cite{tables1954}

\begin{equation}
\begin{aligned}
&S(\omega)
=  \frac{1}{2\pi} \int_{-\infty}^{\infty} 
\langle v(0)v(|t|) \rangle e^{-i\omega t}\,dt
\\& = \frac{N\,k_{B}T}{m} \frac{1}{2\pi}  
\int_{-\infty}^{\infty} e^{-|t|/\tau} e^{-i\omega t}\,dt
\\&= \frac{N\,k_{B}T}{3\pi R \eta}\,\frac{1}{1+(\omega\tau)^{2}}.
\end{aligned}
\tag{15}\label{eq:symmetric_spectrum}
\end{equation}

The real-valued function $S(\omega)$, given by Eq.~(\ref{eq:symmetric_spectrum}), 
 is the power spectral density (PSD) or power spectrum of Brownian motion 
in a memoryless, viscous fluid with shear viscosity~$\eta$. Figure~1 plots, with a heavy dark line the normalized power spectral density given by Eq.~(15) as a function of the dimensionless frequency
  \(\omega\tau = \omega\,\dfrac{m}{6\pi R\,\eta}\).

  More generally, if $\phi(\omega) = \Re_e\{\phi(\omega)\} + i\,\text{Im}\{\phi(\omega)\}$ 
is the Fourier transform of a one-sided time-response function $\psi(t)$, 
 the Fourier transform of the two-sided even function $\psi(|t|)$ is
\begin{equation}
\begin{aligned}
&S(\omega)
=\frac{1}{2\pi} \int_{-\infty}^{\infty} \psi(|t|)\,e^{-i\omega t}\,dt
\\&= 2\frac{1}{2\pi} \int_{0}^{\infty} \psi(t)\cos(\omega t)\,dt 
= 2\,\Re_e\{\phi(\omega)\}.
\end{aligned}
\tag{16}\label{eq:s_omega_relation}
\end{equation}

As an example, the real part of $VAC(\omega)$ given by equation~(\ref{eq:power_spectrum}) is
\begin{equation}
\Re_e\{VAC(\omega)\}
= \frac{N\,k_{B}T}{6\pi k \eta}\,
\frac{1}{1 + (\omega\tau)^{2}}.
\tag{17}\label{eq:real_part_q}
\end{equation}
therefore, $2\,\text{Re}\{VAC(\omega)\} = S(\omega)$, as given by equation~(\ref{eq:symmetric_spectrum}).

  \begin{figure}[t]
    \centering
    \hspace*{-0.2cm}
    \includegraphics[width=\columnwidth]{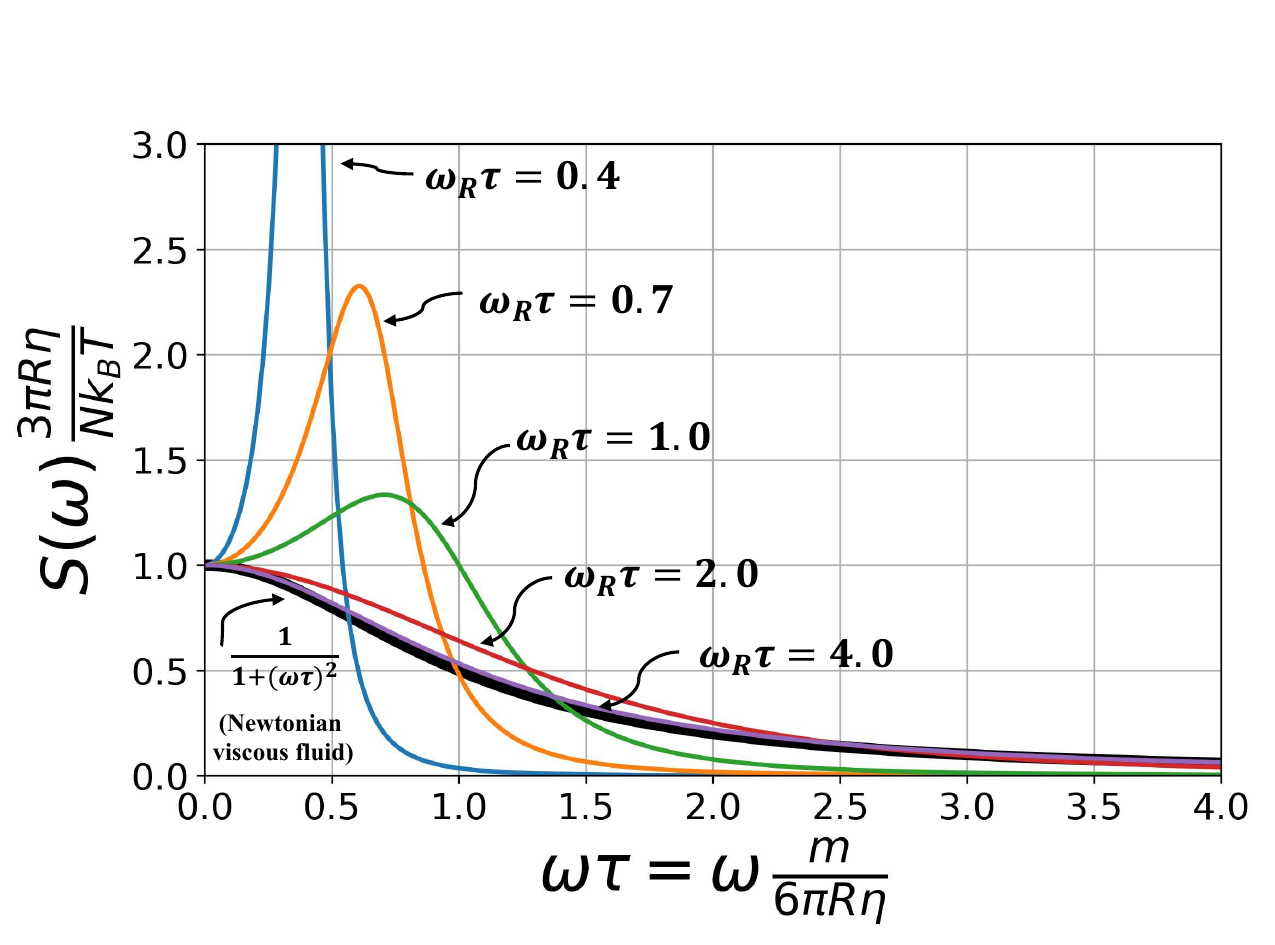}
    \caption{Normalized power spectra of Brownian motion within a Maxwell fluid
with a single relaxation time $\eta/ G$, for different values of the
dimensionless parameter $\omega_R \tau = \frac{1}{\eta}\sqrt{\frac{Gm}{6\pi R}}$
as a function of the dimensionless frequency $\omega \tau = \omega \frac{m}{6\pi R \eta}$.
As the stiffness of the in-series spring increases (large $G$ or large
$\omega_R \tau$), the spectra converge to the power spectrum of Brownian
motion within a memoryless, viscous fluid $(\frac{1}{1+\omega^{2}\tau^{2}})$.}
    \label{fig:fig1_psd}
\end{figure}

\subsection*{Viscous--Viscoelastic Correspondence Principle}

When Brownian particles are immersed in a linear viscoelastic material, the particle motion is described with the generalized Langevin equation~\cite{volkov1984,rodriguez1988}
\begin{equation}
  m\,\frac{d v(t)}{dt}
  \;+\;
  \int_{0}^{t} \zeta (t-\xi)\,v(\xi)\,d\xi
  \;=\;
  f_R(t),
  \tag{18}
\end{equation}
where again $m$ is the mass of the Brownian particle, $v(t)$ is its velocity and $f_R(t)$ is the random force acting on the randomly moving particle. The convolution integral in Eq.~(18) represents the drag force on the particle as it moves randomly within the viscoelastic fluid and accounts for the fading memory of this drag due to the elasticity of the fluid. 
The elastic component of the fluid influences the temporal correlations of the stochastic forces acting on the Brownian particle; therefore, in this case, Eq.~(4) is replaced with
\begin{equation}
  \big\langle f_R(t)\,f_R(0) \big\rangle
  \;=\; k_B T\, \zeta (t-0),
  \tag{19}
\end{equation}
where $ \zeta (t-0)$ is the relaxation kernel of the convolution appearing in the generalized Langevin Eq.~(18).

Mason and Weitz~\cite{mason1995} calculated the mean–square displacement of suspended Brownian particles in the frequency domain by making the assumption that the Stokes result for the drag coefficient on the moving sphere in a memoryless viscous fluid $ \zeta = 6\pi R\eta$~\cite{landau1980fluid}, can be generalized to relate the complex dynamic viscosity of the viscoelastic material, $  \eta_{ve}(\omega)= \frac{\mathcal{G}_{ve}(\omega)}{i\omega}$, to the impedance of the Brownian particle–viscoelastic material network, $ \mathcal{Z}(\omega)
  = \frac{1}{2\pi}
    \int_{-\infty}^{\infty}
      \zeta (t)\, e^{-i\omega t}\, dt,
  \label{eq:Z_of_omega} $
\begin{equation}
  \eta_{ve}(\omega)
  = \frac{\mathcal{G}_{ve}(\omega)}{i\omega}
  = \frac{\mathcal{Z}(\omega)}{6\pi R},
  \tag{20}
\end{equation}

By adopting equation~(20), Mason and Weitz~\cite{mason1995} related the mean-square displacement of the probe Brownian particles,
$\langle \Delta r^{2}(t) \rangle$, to the complex dynamic modulus
$\mathcal{G}_{ve}(\omega)$ of the viscoelastic material within which the particles are suspended.

\begin{equation}
\begin{aligned}
  \langle \Delta r^{2}(\omega) \rangle
 & = \frac{1}{2\pi}
    \int_{}^{}
      \langle \Delta r^{2}(t) \rangle
      e^{-i\omega t}\, dt \\
 & = \frac{N k_{B} T}{3\pi R}
    \frac{1}{i\omega \!\left[\, \mathcal{G}_{ve}(\omega)
      - \dfrac{m}{6\pi R} \omega^{2}\right]},
\end{aligned}
  \tag{21}
\end{equation}

The quantity within the brackets in the denominator of Eq.~(18), $\mathcal{G}_{ve}(\omega)
  - \frac{m}{6\pi R} \omega^{2}= \mathcal{G}_{(\omega)}$, is the complex dynamic modulus of a rheological network that is a parallel connection of the viscoelastic material
within which the Brownian particles are immersed
and an inerter with distributed inertance
$m_R = \dfrac{m}{6\pi R}$~\cite{makris2020}. 

An inerter is a linear mechanical element for which, at the force--displacement level,
the output force is proportional only to the relative acceleration of its end nodes (terminals)~\cite{smith2002,makris2017,makris2018} and complements the set of the three elementary mechanical elements, the other two being the elastic spring and the viscous dashpot.
In a force-current velocity-voltage analogy,
the inerter is the mechanical analog of the electric capacitor and its constant of proportionality is the inertance,
with units of mass~[M].
For instance, a driving spinning top (with a steep lead angle) is a physical realization of an inerter,
since the driving force is only proportional to the relative acceleration of its terminals.
At the stress-strain level, the constant of proportionality of the inerter is the
distributed inertance $m_R$ with units $[M][L]^{-1}$ (i.e.,~Pa·s$^2$).

The inverse of the complex dynamic modulus is the complex dynamic compliance, $\mathcal{J}(\omega)
  = \frac{1}{\mathcal{G}(\omega)}$~\cite{pipkin1986,bird1987,tschoegl1989,giesekus1995,makris2009}
and, $\mathcal{J}(\omega)/i\omega = \mathcal{C}(\omega)$ is the complex creep function--that is the Fourier transform of the creep compliance of the viscoelastic network, $ \mathcal{C}(\omega)
  = \frac{1}{2\pi} \int_{-\infty}^{\infty}
  J (t)\, e^{-i\omega t}\, dt$~\cite{evans2009,makris2019}.
In view of the above relations from the linear theory of viscoelasticity,
the inverse Fourier transform of Eq.~(21) gives~\cite{makris2020}
\begin{equation}
\begin{aligned}
  \langle \Delta r^{2}(t) \rangle
 & = \int_{-\infty}^{\infty}
      \langle \Delta r^{2}(\omega) \rangle\,
      e^{i\omega t}\, d\omega\\
 & = \frac{N k_B T}{3\pi R}\, 
   \int_{-\infty}^{\infty}
       \mathcal{C}(\omega)
      e^{i\omega t}\, d\omega  = \frac{N k_B T}{3\pi R}\,
    J (t),
 \end{aligned}
 \tag{22}
\end{equation}

Equation~(22) is a statement of the viscous--viscoelastic correspondence principle
for Brownian motion, which states that
the mean--square displacement $\langle \Delta r^{2}(t) \rangle$
of a collection of Brownian microspheres with mass~$m$ and radius~$R$
suspended in a linear, isotropic viscoelastic material (fluid or solid) in thermal equilibrium at temperature T,
is identical to $\dfrac{N k_B T}{3\pi R}\, \gamma(t)$,
where $\gamma(t) = J (t)$ is the strain
due to a unit step stress on a rheological network
that is a parallel connection of the linear viscoelastic material
(within which the Brownian microspheres are suspended) and an inerter with distributed inertance
$m_R = \dfrac{m}{6\pi R}$~\cite{makris2020}.
Equation~(22) has an overarching validity regardless what is the mechanism
that the Brownian particles exchange forces with the material within which they are immersed / embedded
(viscous, inertial, hydrodynamic memory, or viscoelastic)~\cite{makris2021,makris2020}.

By replacing in Eq.~(6) the Laplace variable $s$ with $i\omega$ ($s=i\omega$)
and upon substituting the expression of $\langle \Delta r^{2}(\omega) \rangle$
given by Eq.~(21), one obtains
\begin{equation}
\begin{aligned}
  VAC(\omega)
 & = \frac{1}{2\pi}
    \int_{-\infty}^{\infty}
      \langle v(0)\,v(t) \rangle
      e^{-i\omega \tau}\, d\tau
  = -\frac{\omega^{2}}{2}
      \langle \Delta r^{2}(\omega) \rangle \\
 & = \frac{N k_B T}{6\pi R}\,
      \frac{i\omega}{\mathcal{G}(\omega)}
  = \frac{N k_B T}{6\pi R}\,
      \varphi (\omega),
\end{aligned}
  \tag{23}
\end{equation}
The quantity $\frac{i\omega}{\mathcal{G}(\omega)}
  = \varphi (\omega)
  = \frac{\dot{\gamma}(\omega)}{\tau(\omega)}$ is known in rheology as the complex dynamic fluidity~\cite{giesekus1995,makris2009}
and relates a strain--rate output to a stress input.
In structural mechanics the equivalent of
$\varphi (\omega) = i\omega / \mathcal{G}(\omega)$
is the mobility (inverse of the impedance)
and in electrical engineering it is known as the admittance. 

With reference to the reasoning presented following Eq.(13) in association with the result of Eq.(23), the real-valued power spectrum (power spectral density) of Brownian motion
in any linear, isotropic viscoelastic material
that is in thermal equilibrium at temperature~$T$
is given by
\begin{equation}
\begin{aligned}
  \mathcal{S} (\omega)
  = 2\,\Re_e\{\mathrm{VAC}(\omega)\}
 & = \frac{N k_B T}{3\pi R}\,
    \Re_e\!\left\{
      \frac{i\omega}{\mathcal{G}(\omega)}
    \right\} \\
&  = \frac{N k_B T}{3\pi R}\,
   \Re_e\!\left\{ \varphi (\omega) \right\},
 \end{aligned}
 \tag{24}
\end{equation}
where $\varphi (\omega) = \dfrac{i\omega}{\mathcal{G}(\omega)}$
is the complex dynamic fluidity
of a rheological network that is a parallel connection
of the linear viscoelastic material within which the Brownian microparticles are immersed
and an inerter with distributed inertance $m_R = \dfrac{m}{6\pi R}$.

For instance, for Brownian particles immersed in a memoryless viscous fluid
with shear viscosity~$\eta$,
the corresponding rheological network is the \emph{interviscous fluid}
shown in Figure~2~\cite{makris2020,makris2017},
which is a parallel connection of a dashpot
with viscosity~$\eta$ (the viscosity of the solvent)
and an inerter with distributed inertance~$m_R$.

\begin{figure}[t]
    \centering
    \hspace*{-0.5cm}
    \includegraphics[width=\columnwidth]{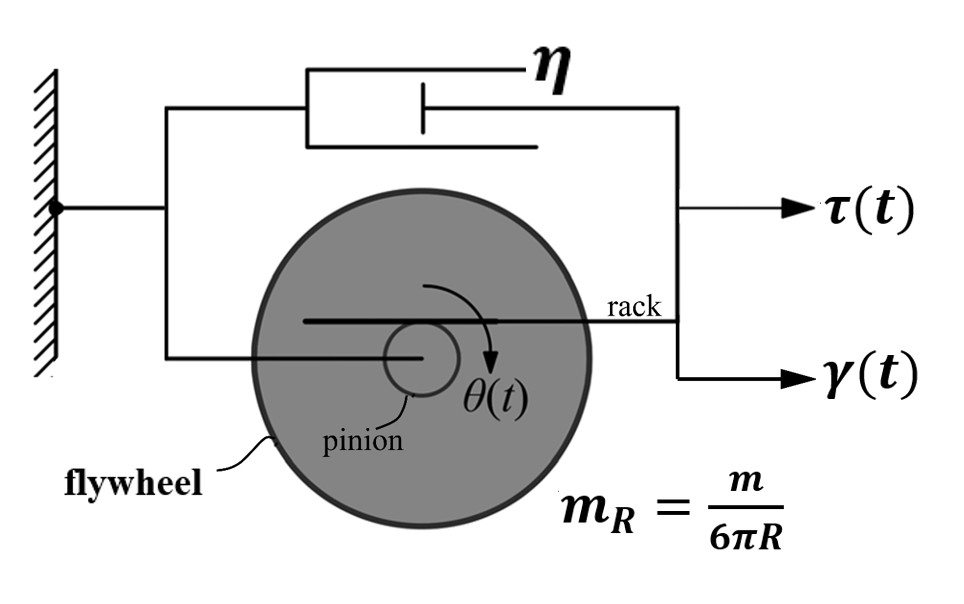}
    \caption{Inertoviscous fluid which is a parallel connection of an inerter
  with distributed inertance $m_R$ with units $[M][L]^{-1}$ and a dashpot with
  viscosity $\eta$ with units $[M][L]^{-1}[T]^{-1}$. In analogy with the
  traditional schematic of a dashpot that is a hydraulic piston, the
  distributed inerter is depicted schematically with a rack--pinion--flywheel
  system.}
    \label{fig:fig2_psd}
\end{figure}

\noindent
Given the parallel connection of the dashpot and the inerter
shown in Figure~2,
the constitutive law of the interviscous fluid is

\begin{equation}
  \tau(t)
  = \eta\, \frac{d\gamma(t)}{dt}
  + m_R\, \frac{d^{2}\gamma(t)}{dt^{2}},
  \tag{25}
\end{equation}

\noindent
The Fourier transform of Eq.~(25) gives
\begin{equation}
  \tau(\omega)
  = \mathcal{G}(\omega)\, \gamma(\omega)
  = \left(i\omega\eta - \omega^{2} m_R\right)\gamma(\omega),
  \tag{26}
\end{equation}
where $\mathcal{G}(\omega) = i\omega\eta - \omega^{2} m_R$
is the complex dynamic modulus of the interviscous fluid. From Eq.~(26),
\begin{equation}
\begin{aligned}
  \varphi (\omega)
  = \frac{i\omega}{\mathcal{G}(\omega)}
  = \frac{i\omega}{i \omega \eta - \omega^{2}m
  _R}
 & = \frac{1}{\eta + i\omega m_R}\\
 & = \frac{1}{\eta}\frac{1}{1 + i\omega\tau}.
\end{aligned}
   \tag{27}
\end{equation}
since $ \frac{m_R}{\eta} = \frac{m}{6\pi R \eta}=\tau $-- that  is the dissipation time of the Brownian process.

Substituting the result of Eq.~(27)
into the expression for the power spectrum
given by Eq.~(24),
the power spectrum for Brownian motion
in a memoryless viscous fluid with viscosity~$\eta$ is
\begin{equation}
  \mathcal{S}(\omega)
  = \frac{N k_B T}{3\pi R}\,
    \mathrm{Re}\{\varphi (\omega) \}
  = \frac{N k_B T}{3\pi R \eta}\,
    \frac{1}{1 + (\omega\tau)^2}
  \tag{28}
\end{equation}
which is the result offered by Eq.~(15)
and plotted in Figure~1 with a heavy dark line.

\subsection*{Power Spectrum of Brownian Motion in a Harmonic Trap}

The Brownian motion of a microparticle embedded
in a viscoelastic Kelvin solid when excited by
the random force $f_{R}(t)$ has been studied by
Uhlenbeck and co-workers~\cite{wang1945,uhlenbeck1930}.
The equation of motion of a microsphere with mass $m$
and radius $R$ in a harmonic trap with viscous damping
subjected to a random excitation force $f_{R}(t)$ is
\begin{equation}
  m\,\frac{d^{2}r(t)}{dt^{2}}
  + \zeta \frac{dr(t)}{dt}
  + k\,r(t)
  = f_{R}(t),
  \tag{29}
\end{equation}
where $r(t)$ is the particle displacement,
$\zeta \frac{dr(t)}{dt}$ is a viscous drag force,
and $k\,r(t)$ is a linear restoring force
proportional to the displacement of the Brownian particle $r(t)$.
Upon dividing Eq.~(29) by the particle mass $m$, one obtains
\begin{equation}
  \frac{d^{2}r(t)}{dt^{2}}
  + \frac{1}{\tau}\,\frac{dr(t)}{dt}
  + \omega_{0}^{2}\,r(t)
  = \frac{f_{R}(t)}{m},
  \tag{30}
\end{equation}
where again $\tau = \frac{m}{\zeta}
       = \frac{m}{6\pi R \eta}$ is the dissipation time and $\omega_{0} = \sqrt{\frac{k}{m}}$ is the undamped natural angular frequency of the trapped particle.
For $\omega_{0} \tau > \tfrac{1}{2}$ the system described by Eq.~(30)
is underdamped, for $\omega_{0} \tau = \tfrac{1}{2}$
the system is critically damped and for $\omega_{0} \tau < \tfrac{1}{2}$ the system is overdamped~\cite{uhlenbeck1930}.

By employing the viscous–viscoelastic correspondence
principle expressed in the time domain by
Eq.~(22) or in the frequency domain by
Eq.~(23) we are interested in evaluating the
complex dynamic fluidity $\varphi (\omega) = \frac{i\omega}{\mathcal{G}(\omega)}$, of the inertoviscoelastic solid shown in Figure~3,
which is a parallel connection of a spring with elastic shear modulus $G$, a dashpot with shear viscosity
$\eta$, and an inerter with distributed inertance
$m_{R}$~\cite{makris2020}.

\begin{figure}[t]
    \centering
    \hspace*{-0.7cm}
    \includegraphics[width=\columnwidth]{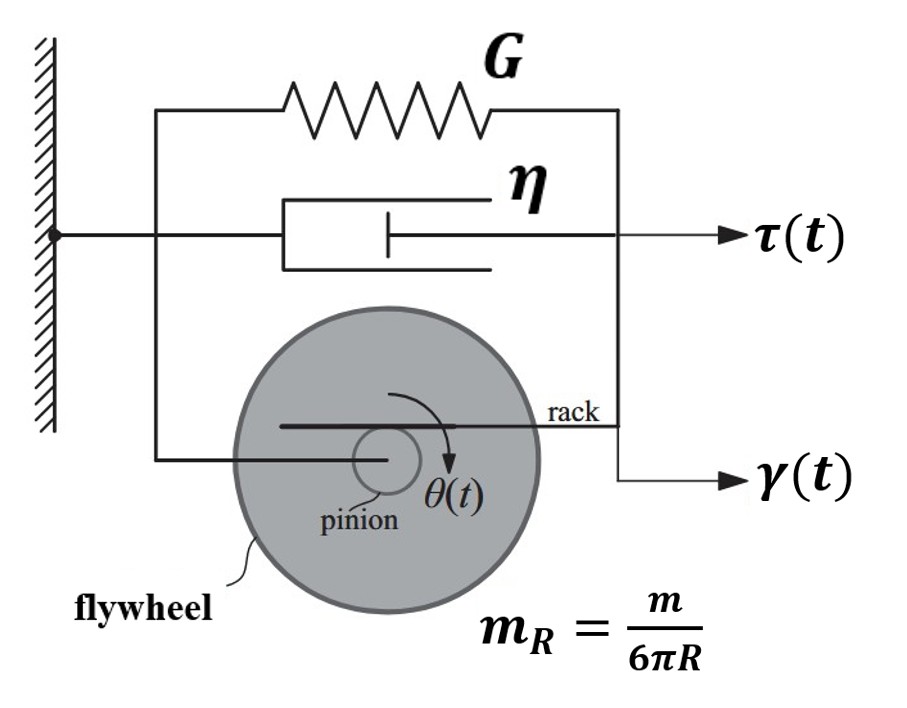}
    \caption{The inertoviscoelastic solid which is a parallel connection of a linear spring 
with elastic shear modulus $G$, a dashpot with shear viscosity $\eta$, 
and an inerter with distributed inertance $m_R$.
}
    \label{fig:fig3_psd}
\end{figure}

Given the parallel connection of the three
elementary rheological idealizations shown in
Figure~3, the constitutive law of the
inertoviscoelastic solid is
\begin{equation}
  \tau(t) = G\,\gamma(t)
           + \eta\,\frac{d\gamma(t)}{dt}
           + m_{R}\,\frac{d^{2}\gamma(t)}{dt^{2}},
  \tag{31}
\end{equation}
The Fourier transform of Eq.~(31) gives
\begin{equation}
  \tau(\omega)
  = \mathcal{G}(\omega)\,\gamma(\omega)
  = \bigl(G + i\omega\eta - \omega^{2}m_{R}\bigr)
    \gamma(\omega).
  \tag{32}
\end{equation}

\noindent
where $\mathcal{G} (\omega) = G + i \omega \eta - \omega^2 m_R$ is the complex dynamic modulus of the inertoviscoelastic solid shown in Figure~3.
Accordingly, the complex dynamic fluidity of the inertoviscoelastic solid is

\begin{equation}
\begin{aligned}
\varphi (\omega) = \frac{i\omega}{\mathcal{G} (\omega)} 
&= \frac{i\omega}{G + i\omega \eta - \omega^2 m_R} \\
&= \frac{1}{m_R} \cdot
\frac{\omega^2/\tau \, + i \omega (\omega_R^2 - \omega^2)}{(\omega_R^2 - \omega^2)^2 + \left( \frac{\omega}{\tau} \right)^2}.
\end{aligned}
\tag{33}
\end{equation}

\noindent
where $\omega_R = \sqrt{\frac{G}{m_R}} = \sqrt{\frac{6\pi R\, G}{m}} = \sqrt{\frac{k}{m}}= \omega_0$ is the undamped angular frequency, $m_R= \frac{m}{6 \pi R}$ 
and $\tau = \frac{m}{6\pi R \eta} = \frac{m_R}{\eta}$ is the dissipation time of the Brownian process.

Substitution of the expression of the complex dynamic fluidity \(\varphi(\omega)\)
given by Eq.~(33) into the general expression for the power spectral density,
given by Eq.~(24), gives:

\begin{equation}
\begin{aligned}
\mathcal{S}(\omega) 
&= \frac{N k_B T}{3\pi R }
\Re_e\{ \varphi (\omega) \}\\
&= \frac{N k_B T}{3\pi R \eta}
\frac{(\omega\tau)^2}{\left[(\omega_R\tau)^2 - (\omega\tau)^2\right]^2 + (\omega\tau)^2}.
\end{aligned}
\tag{34}
\end{equation}

\medskip
Figure~4 plots the normalized power spectrum (PSD) of the Brownian motion 
of a particle trapped in a harmonic potential given by equation~(34)
as a function of the dimensionless frequency 
\(\omega\tau = \omega \frac{m}{6\pi R \eta}\),
for different values of the dimensionless network parameter $\omega_R \tau = \sqrt{\frac{6 \pi R G}{m}} \tau=  \frac{1}{\eta} \sqrt{\frac{Gm}{6\pi R}}.$

\begin{figure}[t]
    \centering
    \hspace*{-0.3cm}
    \includegraphics[width=1.15\columnwidth]{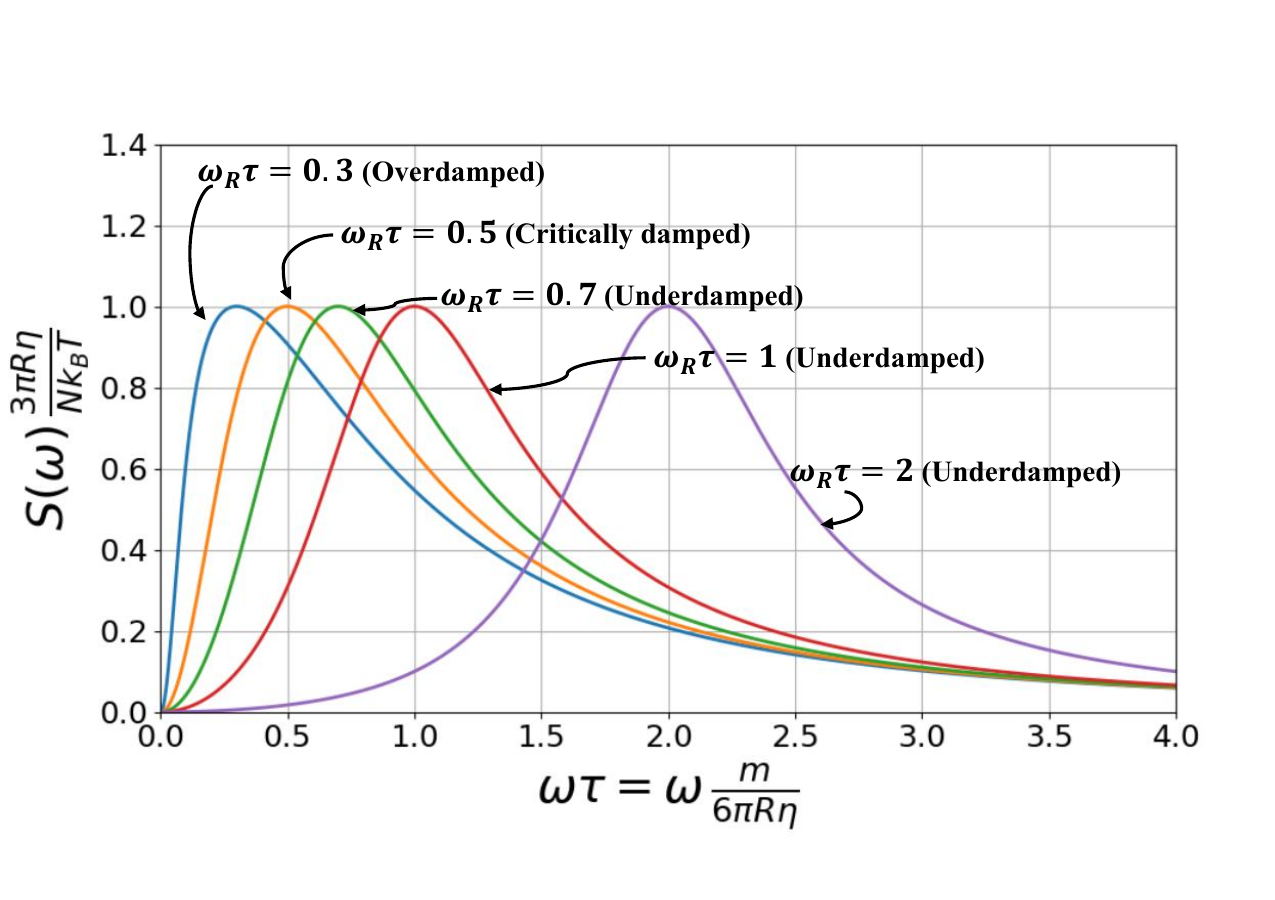}
    \caption{Normalized power spectra of Brownian motion of particles suspended
in a Kelvin solid (harmonic trap) for different values of the
dimensionless parameter $\omega_R \tau = \frac{1}{\eta}\sqrt{\frac{G m}{6\pi R}}$
as a function of the dimensionless frequency
$\omega \tau = \omega \frac{m}{6\pi R \eta}$.
}
    \label{fig:fig4_psd}
\end{figure}

\section*{Power Spectrum of Brownian Motion within a Maxwell Fluid}

The Brownian motion of particles immersed in a Maxwell fluid with a single 
relaxation time $\frac{\eta}{G}$, when subjected to the random force $f_R(t)$ 
from the collisions of the molecules of the viscoelastic fluid is described with the Langevin equation. (18), where the relaxation kernel $\zeta (t-\xi)$ is~\cite{volkov1984,rodriguez1988}
\[
\zeta (t-\xi) = 6\pi R\, G_{ve}(t-\xi)
= 6\pi R\, G\, e^{-\frac{G}{\eta}(t-\xi)} . \tag{35}
\]
Here, $G_{ve}(t-\xi) = G e^{-\frac{G}{\eta}(t-\xi)}$ is the relaxation modulus [stress due to a unit–amplitude step strain 
$\gamma(t) = U(t-0)$] of the Maxwell fluid~\cite{pipkin1986,bird1987,tschoegl1989,giesekus1995,makris2009}. 
Equation (18), in association with Eq.~(35), leads to the temporal 
evolution of the Brownian particle's velocity autocorrelation function~\cite{volkov1996}, from which the power spectrum 
$S(\omega)$ can, in principle, be computed with the application of 
integral transforms.

In this section, the power spectrum of Brownian motion within a Maxwell 
fluid is calculated by using equation~(24), which derives from the 
viscous–viscoelastic correspondence principle as expressed by equation (21) in the frequency domain. 
Accordingly, the problem reduces to the calculation of the complex dynamic 
fluidity, $\varphi (\omega) = \frac{i\omega}{\mathcal{G}(\omega)},$ of a Maxwell fluid with shear modulus $G$ and shear viscosity $\eta$, 
that is connected in parallel with an inerter with a distributed 
inertance $m_R = \frac{m}{6\pi R},$ as shown in Fig.~5.

\begin{figure}[t]
    \centering
    \hspace*{-0.5cm}
    \includegraphics[width=\columnwidth]{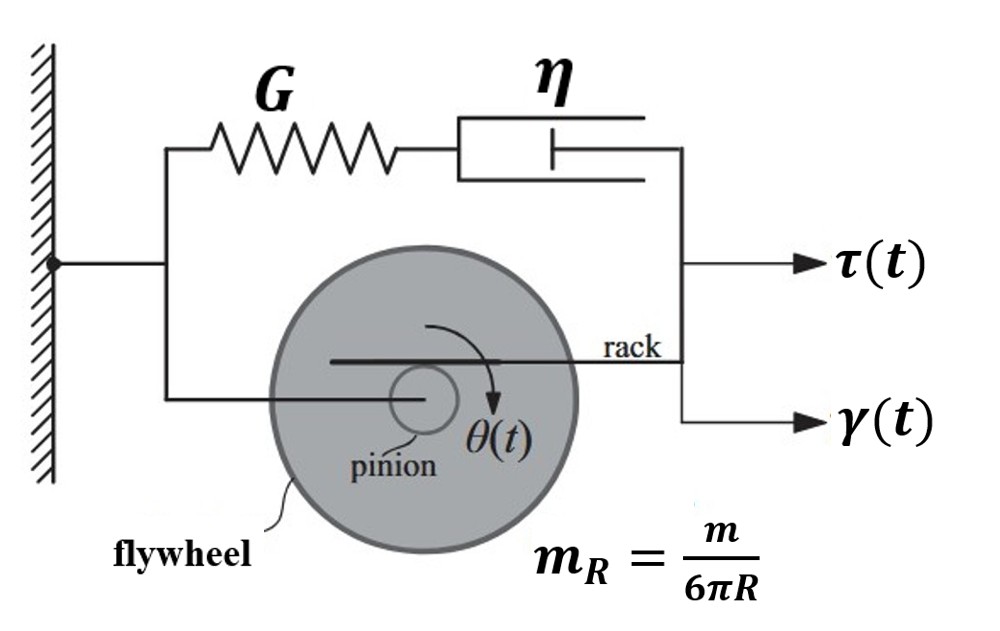}
    \caption{Rheological analogue for Brownian motion in a Maxwell fluid.
It consists of the Maxwell element with shear modulus $G$ and shear
viscosity $\eta$ that is connected in parallel with an inerter with
distributed inertance $m_R = \frac{m}{6\pi R}$.}
    \label{fig:fig5_psd}
\end{figure}

The total stress output, $\tau(t)=\tau_1(t)+\tau_2(t)$ from the rheological network shown in Fig.~5 is the 
summation of the stress output from the Maxwell element $\tau_1(t)$
\[
\tau_1(t) + \frac{\eta}{G}\,\frac{d\tau_1(t)}{dt}
= \eta\,\frac{d\gamma(t)}{dt},
\tag{36}
\]
and the stress output from the inerter $\tau_2(t)$
\[
\tau_2(t) = m_R\,\frac{d^{2}\gamma(t)}{dt^{2}}.
\tag{37}
\]
The summation of Eqs.~(36) and (37), together with the time derivative of 
Eq.~(37), yields a third–order constitutive equation for the rheological 
network shown in Figure~5.

\[
\tau(t)
+ \frac{\eta}{G}\,\frac{d\tau(t)}{dt}
= \eta\,\frac{d\gamma(t)}{dt}
+ m_R\,\frac{d^{2}\gamma(t)}{dt^{2}}
+ \frac{\eta m_R}{G}\,\frac{d^{3}\gamma(t)}{dt^{3}},
\tag{38}
\]
By defining the dissipation time $\tau = \frac{m_R}{\eta}=\frac{m}{6 \pi R \eta}$ and the rotational angular frequency \( \omega_R = \sqrt{\frac{G}{m_R}}
= \sqrt{\frac{6\pi R G}{m}} \), as in the previous case, equation~(38) assumes the form
\[
\tau(t)
+ \frac{1}{\tau \omega_R^{2}}\,
\frac{d\tau(t)}{dt}
=
m_R\left(
\frac{1}{\tau}\frac{d\gamma(t)}{dt}
+ \frac{d^{2}\gamma(t)}{dt^{2}}
+ \frac{1}{\tau \omega_R^{2}}\frac{d^{3}\gamma(t)}{dt^{3}}
\right).
\tag{39}
\]

The Fourier transform of Eq.~(39) gives $\tau(\omega)
= \mathcal{G} (\omega)\,\gamma(\omega),$ where \(\mathcal{G} (\omega)\) is the complex dynamic modulus of the rheological network 
shown in Fig.~5:

\[
\mathcal{G} (\omega)
= \dfrac{\tau(\omega)}{\gamma(\omega)} = m_R\,\frac{
i\omega\left(\frac{1}{\tau}+i\omega-\frac{1}{\tau\omega_R^{2}}\,\omega^{2}\right)
}{
1 + i\omega \frac{1}{\tau\omega_R^{2}}
}
\tag{40}
\]

Accordingly, the complex dynamic fluidity of the rheological model described by Eq.~(39) is
\begin{equation}
\begin{aligned}
&\varphi(\omega)
= \frac{i\omega}{\mathcal{G}(\omega)}
= \frac{1}{m_R}\,
\frac{
1 + i\omega\,\dfrac{1}{\tau\omega_R^{2}}
}{
\dfrac{1}{\tau} + i\omega - \dfrac{1}{\tau\omega_R^{2}}\omega^{2}
}
\\&= \frac{1}{m_R}\,
\frac{
\tau\omega_R^{4} + i\omega(\omega_R^{2}-\omega^{2})
}{
(\omega_R^{2}-\omega^{2})^{2} + (\omega\tau)^{2}\omega_R^{4}
}
\end{aligned}
\tag{41}
\end{equation}

Substitution of the real part of the expression of the complex dynamic fluidity $\varphi (\omega)$ 
given by equation~(41) in the general expression for the power spectral density 
given by Eq.~(24) gives
\begin{equation}
\begin{aligned}
&\mathcal{S} (\omega)
= \frac{N k_B T}{3\pi R}\,
\Re_e\{\varphi(\omega)\}
\\&= \frac{N k_B T}{3\pi R \eta}\,
\frac{(\omega_R\tau)^4}{
\left[(\omega_R \tau)^{2}-(\omega \tau)^{2}) \right]^{2} 
+ (\omega\tau)^{2}(\omega_R \tau)^{4}}
\end{aligned}
\tag{42}
\end{equation}

Figure~1 plots the normalized power spectrum (PSD) of Brownian motion 
of particles immersed in a Maxwell fluid, given by Eq.~(42), as a function 
of the dimensionless frequency $\omega\tau = \omega \frac {m}{6 \pi R  \eta}$, for different 
values of the dimensionless network parameter $\omega_R \tau 
= \frac{1}{\eta}\sqrt{\frac{G m}{6\pi R}}.$ As the value of $\omega_R\tau$ increases (large $G$ or small $\eta$), 
the power spectra tend to the power spectrum for Brownian motion in
a memoryless viscous fluid. plotted with a heavy dark line.

\vspace{1\baselineskip}

\textbf{Power Spectrum of Brownian Motion within a Jeffreys Fluid.}

When a Maxwell element (a spring $G$ and a dashpot $\eta$ connected in
series) is connected in parallel with a dashpot with viscosity
$\eta_{\infty}$, it forms a rheological network known as the Jeffreys
fluid, which has been proposed by Jeffreys~\cite{jeffreys1929} to model the
viscoelastic behavior of earth strata. The Jeffreys fluid has enjoyed wide application by rheologists in studies
ranging from the onset of convection in viscoelastic fluids~\cite{lebon1994}, the description of the linear response of selected soft materials such as wormlike micellar solutions and concentrated dispersions~\cite{khan2014soft,khan2014pre}, to the understanding of peristaltic transport~\cite{kothandapani2008}. At very low
frequencies there is a slow relaxation typically arising from the
reorganization of the colloidal structure in the viscoelastic material with
relaxation time $\dfrac{\eta}{G}$. At high frequencies, because of the compliant
spring $G$, the shear stresses are primarily resisted by the parallel
dashpot with viscosity $\eta_{\infty}$, and the response is viscously dominated.
Accordingly, the relaxation modulus $G_{\mathrm{ve}}(t)$ of the Jeffreys
fluid is~\cite{bird1987,makris2009,makris2021N}
\[
G_{\mathrm{ve}}(t) = \eta_{\infty}\,\delta(t-0) + G\,e^{-\,\frac{G}{\eta}t}.
\tag{43}
\]

By employing the viscous--viscoelastic correspondence
principle expressed in the time domain by equation~(22)
or in the frequency domain by equation~(23), the Brownian
motion of particles immersed in a Jeffreys fluid can be
analysed by computing the time and frequency response
functions of the rheological network shown in Figure~6.
The total stress $\tau(t)=\tau_1(t)+\tau_2(t)+\tau_3(t)$ from the
rheological network shown in Figure~6 is the summation of the stress
output from the Maxwell element shown on the top of Figure~6,
\begin{equation}
\tau_1(t) + \frac{\eta}{G}\,\frac{d\tau_1(t)}{dt}
   = \eta\,\frac{d{\gamma}(t)}{dt}.
\tag{44}
\end{equation}
the stress output from the dashpot with viscosity $\eta_{\infty}$,
\begin{equation}
\tau_2(t) = \eta_{\infty}\,\frac{d\gamma(t)}{dt}.
\tag{45}
\end{equation}
and the stress output from the inerter shown at the bottom of Figure~6
\begin{equation}
\tau_3(t) = m_R\,\frac{d^{2}\gamma(t)}{dt^{2}}.
\tag{46}
\end{equation}

\begin{figure}[t]
    \centering
    \includegraphics[width=\columnwidth]{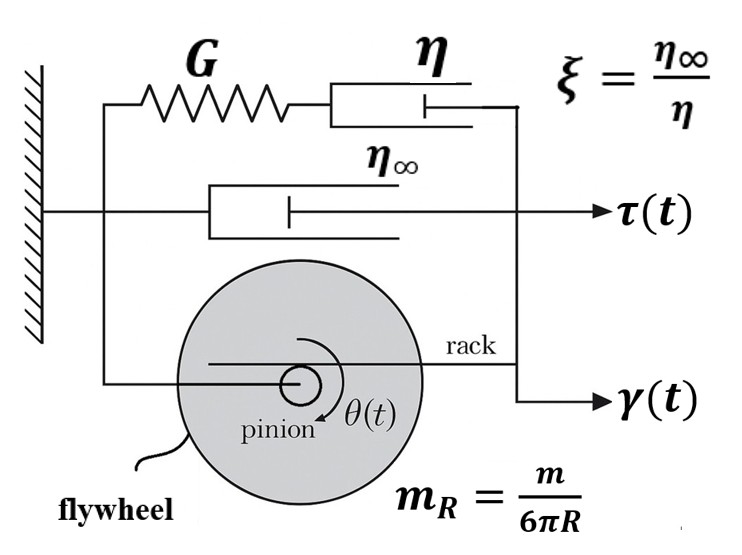}
    \caption{Rheological analogue for Brownian motion of microspheres with
    mass $m$ and radius $R$ immersed in a Jeffreys fluid described with
    a Maxwell element--dashpot parallel connection.}
    \label{fig:jeffreys_rheology}
\end{figure}

The summation of equations (44)--(46) together with the time derivatives
of equations (45) and (46) yields a third-order constitutive equation
for the rheological network shown in Fig.~6~\cite{makris2021N}:

\begin{equation}
\begin{aligned}
\tau(t) + \frac{\eta}{G}\frac{d\tau(t)}{dt}
&= m_R \Bigg[
\frac{\eta}{m_R}\!\left(1+\frac{\eta_{\infty}}{\eta}\right)\frac{d\gamma(t)}{dt}\\
&\quad
+ \left(1+\frac{\eta. \eta_{\infty}}{m_R G}\right)\frac{d^{2}\gamma(t)}{dt^{2}}
+ \frac{\eta}{G}\frac{d^{3}\gamma(t)}{dt^{3}}
\end{aligned}
\tag{47}
\end{equation}
By defining the dissipation time 
$\tau = \dfrac{m_R}{\eta} = \dfrac{m}{6\pi R\eta}$,
the rotational angular frequency
$\omega_R = \sqrt{\dfrac{G}{m_R}} = \sqrt{\dfrac{6\pi R G}{m}}$,
and the dimensionless viscosity ratio
$\xi = \dfrac{\eta_{\infty}}{\eta}$,
equation~(47) assumes the form

\begin{equation}
\begin{aligned}
\tau(t) + \frac{1}{\tau \omega_R^{2}}\frac{d\tau(t)}{dt}
&= m_R \bigg[
\frac{1}{\tau}(1+\xi)\frac{d\gamma(t)}{dt}\\
&+ \left(1 + \frac{\xi}{(\omega_R \tau)^{2}}\right)
\frac{d^{2}\gamma(t)}{dt^{2}}
+ \frac{1}{\tau \omega_R^{2}}\frac{d^{3}\gamma(t)}{dt^{3}}
\bigg].
\end{aligned}
\tag{48}
\end{equation}

Equation~(48) is of the same form as equation~(39); however, now the 
coefficients of the first and second time--derivatives of the shear strain 
contain the viscosity ratio $\xi=\eta_{\infty}/\eta$, which controls the 
effects of the in--parallel dashpot whose viscosity $\eta_{\infty}$ becomes 
dominant at high frequencies.

The Fourier transform of equation~(48) gives $\tau(\omega)=G(\omega)\,\gamma(\omega),$
where $G(\omega)$ is the complex dynamic modulus of the rheological 
network shown in Figure~6.

\begin{equation}
G(\omega)
= \frac{\tau(\omega)}{\gamma(\omega)}
= m_R\, i\omega\,
\frac{
\frac{1+\xi}{\tau}
+ i\omega\left(1+\frac{\xi}{(\omega_R\tau)^2}\right)
- \frac{\omega^2}{\tau\omega_R^2}
}{
1 + \frac{i\omega}{\tau\omega_R^2}
}.
\tag{49}
\end{equation}

Accordingly, the complex dynamic fluidity is
\begin{equation}
\begin{aligned}
\varphi(\omega)
&= \frac{i\omega}{G(\omega)}\\
&= \frac{1}{m_R}
\frac{
\tau\omega_R^{2} + i\omega
}{
(1+\xi)\,\omega_R^{2}
- \omega^{2}
+ i\omega\left(1+\frac{\xi}{(\omega_R\tau)^{2}}\right)\tau\omega_R^{2}
}.
\end{aligned}
\tag{50}
\end{equation}

Substitution of the real part of the expression of the complex dynamic
fluidity $\varphi(\omega)$ given by equation~(50) in the general expression
for the power spectral density given by Eq.~(24) gives

\begin{widetext}
\begin{equation}
S(\omega)
= \frac{N k_B T}{3\pi R}\,\Re_e\{\varphi(\omega)\}
= \frac{N k_B T}{3\pi R \eta}
\frac{
(\omega_R\tau)^{2}
\left[(1+\xi)(\omega_R\tau)^{2}
+ \frac{\xi}{(\omega_R\tau)^{2}}(\omega\tau)^{2}\right]
}{
\left[(1+\xi)(\omega_R\tau)^{2} - (\omega\tau)^{2}\right]^{2}
+ \left[\omega\tau\left(1 + \frac{\xi}{(\omega_R\tau)^{2}}\right)
(\omega_R\tau)^{2}\right]^{2}}
\tag{51}
\end{equation}
\end{widetext}

Figure~7 plots the normalized power spectrum (PSD) of Brownian motion of
particles immersed in a Jeffreys fluid given by equation~(51) as a function
of the dimensionless frequency $\omega\tau =\omega \dfrac{m}{6\pi R\eta}$ for
different values of the dimensionless parameters of the rheological
network $\omega_R\tau = \dfrac{1}{\eta}\sqrt{\frac{Gm}{6\pi R}}$ and
$\xi = \dfrac{\eta_{\infty}}{\eta}$.
For $\xi = 0$ we recover the power spectra for Brownian motion of
particles immersed in a Maxwell fluid shown in Figure~1.

\begin{figure}[t]
    \centering
    \includegraphics[width=\columnwidth]{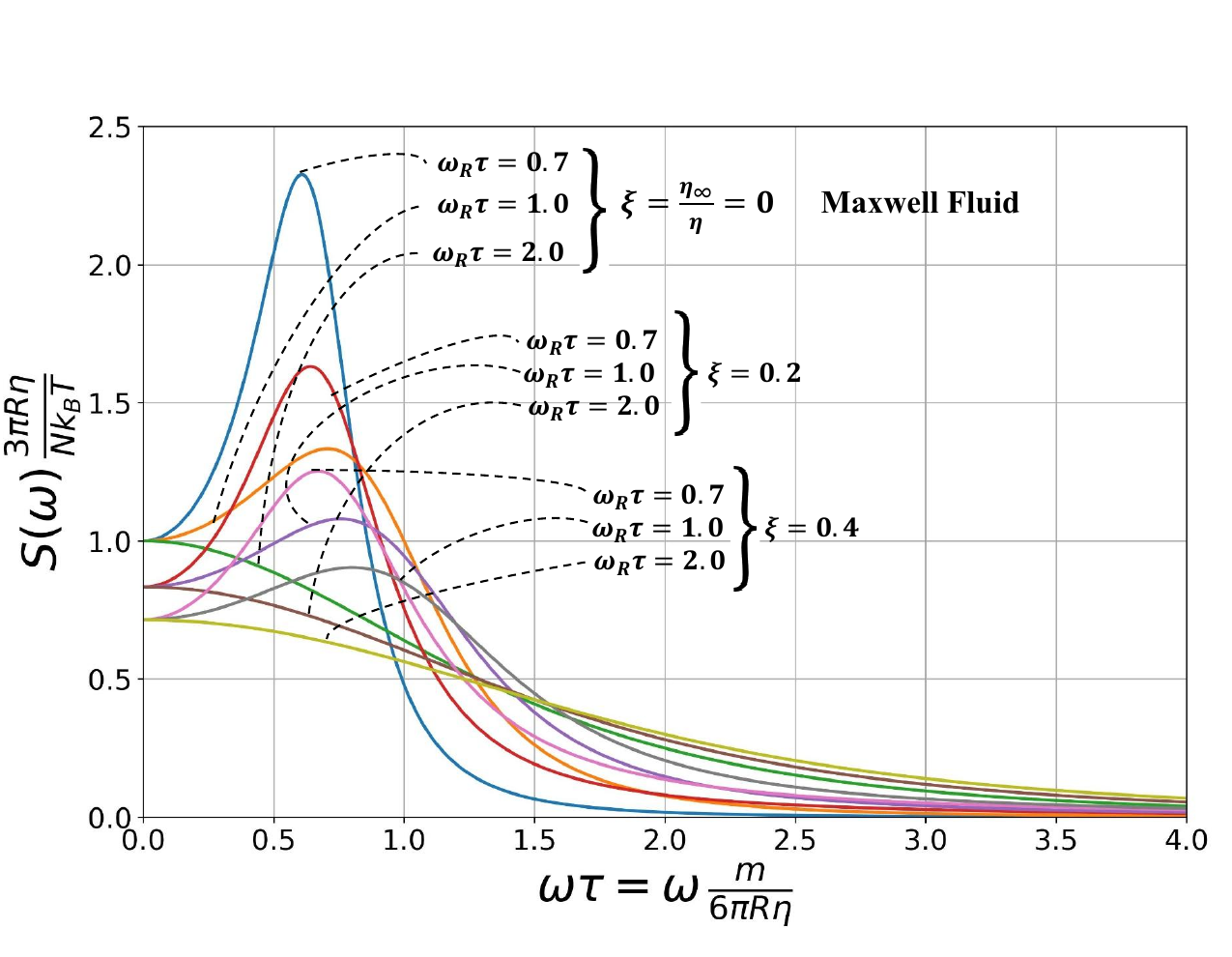}
    \caption{Normalized power spectra of Brownian motion within a Jeffreys fluid for various 
values of the dimensionless parameters $\omega_R\tau = \dfrac{1}{\eta}\sqrt{\frac{Gm}{6\pi R}}$ and $\xi = \frac{\eta_\infty}{\eta}$, shown as a 
function of the dimensionless frequency $\omega \tau = \omega m/(6\pi R\eta)$}
\end{figure}

\section*{Power Spectrum of Brownian Motion Within a Subdiffusive Material}

Several complex materials exhibit a subdiffusive behavior where from
early times and over several temporal decades the mean--square
displacement grows with time according to a power law
$\langle \Delta r^{2}(t) \rangle \sim t^{\alpha}$, where $0 \le \alpha \le 1$
is the diffusive exponent~\cite{palmer1998,gisler1999}.

This type of power–law rheological behavior was first reported by Nutting~\cite{nutting1921}, who noticed that the stress response of
several fluidlike materials to a step strain decays following a power law $\tau(t) = G_{\mathrm{ve}}(t) \sim t^{-\alpha},$ with $ 0\le \alpha \le 1.$ Following Nutting's observations and the early work of Gemant~\cite{gemant1936,gemant1938}
on fractional differentials, Scott Blair~\cite{scottblair1944,scottblair1947} pioneered the introduction of fractional calculus in viscoelasticity. In analogy
to the Hookean spring, in which the stress is proportional to the zeroth
derivative of the strain, and the Newtonian dashpot, in which the stress
is proportional to the first derivative of the strain, Scott Blair and
co-workers~\cite{scottblair1944,scottblair1947,scottblair1949} proposed the
springpot element, which is a mechanical element between a spring and a
dashpot with constitutive law
\begin{equation}
\tau(t) = \mu_{\alpha}\, \frac{d^{\alpha}\gamma(t)}{dt^{\alpha}},
\qquad 0 \le \alpha \le 1,
\tag{52}
\end{equation}
where $\alpha$ is a positive real number ($0 \le \alpha \le 1$),
$\mu_{\alpha}$ is a phenomenological material parameter with units
$[M][L]^{-1}[T]^{\alpha-2}$ (i.e.\ Pa·s$^{\alpha}$), and
$\dfrac{d^{\alpha}\gamma(t)}{dt^{\alpha}}$ is the fractional derivative of order
$\alpha$ of the strain history $\gamma(t)$.

A definition of the fractional derivative of order $\alpha$ is given by
the convolution integral
\begin{equation}
I^{\alpha}\gamma(t) =
\frac{1}{\Gamma(\alpha)}
\int_{c}^{t} (t-\xi)^{\alpha-1}\,\gamma(\xi)\,d\xi.
\tag{53}
\end{equation}
where $\Gamma(\alpha)$ is the Gamma function. When the lower limit
$c=0$, the integral given by Eq.~(53) is often referred to as the
Riemann--Liouville fractional integral~\cite{oldham1974,samko1974,miller1993,podlubny1998}.
The integral in Eq.~(53) converges only for $\alpha>0$, or in the case
where $\alpha$ is a complex number, the integral converges for
$\Re_e(\alpha)>0$. Nevertheless, by a proper analytic continuation across
the line $\Re_e(\alpha)=0$ and provided that the function $\gamma(t)$ is
$n$ times differentiable, it can be shown that the integral given by
Eq.~(58) exists for $n-\Re_e(\alpha)>0$~\cite{riesz1949}. In this case the
fractional derivative of order $\alpha\in\mathbb{R}^{+}$ exists and is
defined as
\begin{equation}
\frac{d^{\alpha}\gamma(t)}{dt^{\alpha}}
= I^{-\alpha}\gamma(t)
= \frac{1}{\Gamma(-\alpha)}
\int_{0^{-}}^{t} \frac{\gamma(\xi)}{(t-\xi)^{\alpha+1}}\, d\xi,
\qquad \alpha\in\mathbb{R}^{+},
\tag{54}
\end{equation}
where $\mathbb{R}^{+}$ is the set of positive real numbers, and the lower
limit of integration $0^{-}$ may engage an entire singular function at
the origin such as $\gamma(t)=\delta(t-0)$~[13]. Equation~(54)
indicates that the fractional derivative of order $\alpha$ of
$\gamma(t)$ is essentially the convolution of $\gamma(t)$ with the
kernel $\frac{t^{-\alpha-1}}{\Gamma(-\alpha)}$~\cite{oldham1974,miller1993,mainardi2010}. The Riemann--Liouville
definition of the fractional derivative of order
$\alpha\in\mathbb{R}^{+}$ given by Eq.~(54), where the lower limit of
integration is zero, is relevant to rheology since the strain and stress
histories $\gamma(t)$ and $\tau(t)$ are causal functions, being zero at
negative times.

The relaxation modulus (stress history due to a unit-amplitude step
strain $\gamma(t)=U(t-0)$) of the springpot element (Scott Blair fluid)
expressed by Eq.~(52) is~\cite{smit1970,koeller1984,friedrich1991,heymans1994}
\begin{equation}
G_{ve}(t)
= \mu_{\alpha}\, \frac{1}{\Gamma(1-\alpha)}\, t^{-\alpha},
\qquad t>0,
\tag{55}
\end{equation}
which decays by following the power law initially observed by
Nutting~\cite{nutting1921}. The creep compliance (retardation function) of the
springpot element is~\cite{mainardi2010,smit1970,koeller1984,friedrich1991,heymans1994,schiessel1995,palade1996,makris2020E}

\begin{equation}
J_{ve}(t)
= \frac{1}{\mu_{\alpha}}\, \frac{1}{\Gamma(1+\alpha)}\, t^{\alpha},
\qquad t \ge 0.
\tag{56}
\end{equation}

The power law $t^{\alpha}$ appearing in Eq.~(56) renders the elementary
springpot element expressed by Eq.~(52) (Scott Blair fluid) a suitable
phenomenological model to study Brownian motion in subdiffusive
materials.

By employing the viscous--viscoelastic correspondence principle
expressed in the time domain by equation~(22), the Brownian motion of
particles immersed in a subdiffusive material can be analysed by
computing the frequency and time response function of the rheological
network shown in Figure~8, which is a parallel connection of the
springpot element described by equation~(52) and an inerter with
distributed inertance $m_R = \frac{m}{6\pi R}$

\begin{figure}[t]
    \centering
    \includegraphics[width=\columnwidth]{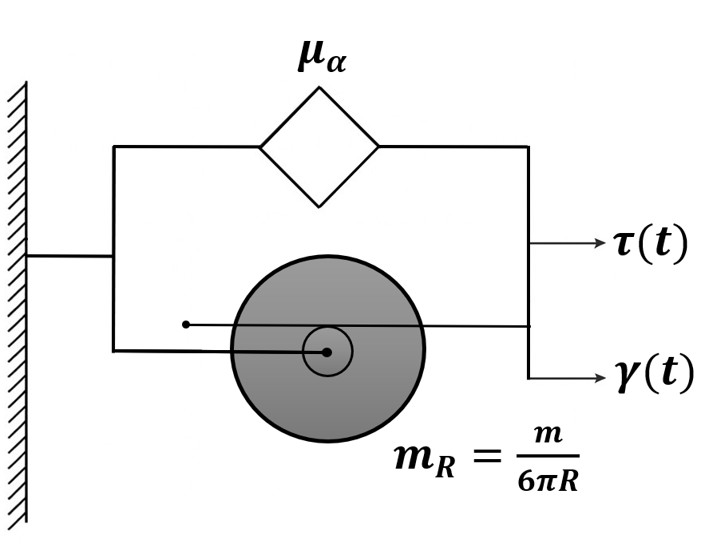}
    \caption{Rheological analogue for Brownian motion in a subdiffusive
    material. It consists of a springpot with phenomenological
    material parameter $\mu_{\alpha}$ $( 0\le \alpha \le 1)$ that is connected in parallel with an
    inerter with distributed inertance $m_R = \frac{m}{6\pi R}$.}
    \label{fig:fig8_springpot_inerter}
\end{figure}

Given the parallel connection of the springpot and the 
inerter the constitutive 
law of the rheological network shown in Fig.~\ref{fig:fig8_springpot_inerter} is

\begin{equation}
    \tau(t) = 
    \mu_{\alpha}\,\frac{d^{\alpha}\gamma(t)}{dt^{\alpha}}
    + 
    m_R\,\frac{d^{2}\gamma(t)}{dt^{2}},
\tag{57}
\end{equation}
The Fourier transform of Equation~(57) gives
\begin{equation}
    \tau(\omega)
    =
    G(\omega)\,\gamma(\omega)
    =
    \left[(i\omega)^{\alpha}\mu_{\alpha}
    -\omega^{2} m_R\right]\gamma(\omega),
\tag{58}
\end{equation}

\par\noindent
where $(i\omega)^{\alpha}=\omega^{\alpha}\left(\cos\frac{\alpha\pi}{2}
+ i\sin\frac{\alpha\pi}{2}\right)$. The quantity in brackets on the
right-hand side of Eq.~[58] is the complex dynamic modulus
of the rheological network shown in Fig.~8.

\begin{equation}
G(\omega)
= \frac{\tau(\omega)}{\gamma(\omega)}
= \mu_{\alpha}\,\omega^{\alpha}
\left(
    \cos\frac{\alpha\pi}{2}
    + i \sin\frac{\alpha\pi}{2}
\right)
- m_{R}\,\omega^{2}.
\tag{59}
\end{equation}

The complex dynamic fluidity of the rheological model described by equation~(57) and schematically shown in Figure~8 is

\begin{equation}
\begin{aligned}
\varphi(\omega)
&= \frac{i\omega}{G(\omega)}\\
&=\dfrac{1}{m_R} \dfrac{i\omega}{\dfrac{\mu_{\alpha}}{m_{R}}\,
\omega^{\alpha} 
\cos\frac{\alpha\pi}{2}
- \omega^{2}+i\,\dfrac{\mu_{\alpha}}{m_{R}}\,\omega^{\alpha}\sin\!\left(\frac{\alpha\pi}{2}\right)
}
\end{aligned}
\tag{60}
\end{equation}

By recognizing that the quantity $\lambda = \left(\frac{m_{R}}{\mu_{\alpha}}\right)^{\frac{1}{2-\alpha}}$ has units of time, the real part of the complex dynamic fluidity $\varphi(\omega)$ given by equation~(60) assumes the form
\begin{widetext}
\begin{equation}
\Re_e\{\varphi(\omega)\}
=
\frac{\lambda}{m_{R}}
\frac{
\lambda^{2-\alpha} \, \dfrac{\mu_{\alpha}}{m_{R}} (\omega \lambda)^{\alpha+1} \, 
\sin\left(\frac{\alpha \pi}{2}\right)
}{
\Big[
\lambda^{2-\alpha}
\dfrac{\mu_{\alpha}}{m_{R}}
(\omega \lambda)^{\alpha}
\cos\left(\frac{\alpha \pi}{2}\right)
-
(\omega \lambda)^{2}
\Big]^{2}
+
\Big[
\lambda^{2-\alpha}
\dfrac{\mu_{\alpha}}{m_{R}}
(\omega \lambda)^{\alpha}
\sin\left(\frac{\alpha \pi}{2}\right)
\Big]^{2}
}
\tag{61}
\end{equation}
\end{widetext}

Upon using that $\lambda^{\,2-\alpha}\frac{\mu_{\alpha}}{m_{R}} = 1,$ the substitution of equation (61) in the general expression for the
power spectral density given by Eq.~(24) gives
\begin{widetext}
\begin{equation}
S(\omega)
=
\frac{N k_{B} T}{3 \pi R}\,
\Re\{ \varphi(\omega)\}
=
\frac{N k_{B} T}{3 \pi R}\,
\frac{1}{(\mu_{\alpha}\, m_{R}^{\,1-\alpha})^{\frac{1}{\,2-\alpha\,}}}
\frac{
(\omega \lambda)^{1+\alpha}\,
\sin\left(\frac{\alpha\pi}{2}\right)
}{
\left[
(\omega\lambda)^{\alpha}
\cos\left(\frac{\alpha\pi}{2}\right)
-(\omega\lambda)^{2}
\right]^{2}
+
\left[
(\omega\lambda)^{\alpha}\,
\sin\left(\frac{\alpha\pi}{2}\right)
\right]^{2}
}.
\tag{62}
\end{equation}
\end{widetext}

where the quantity 
\(
\big[\mu_{\alpha} m_{R}^{\,1-\alpha}\big]^{\tfrac{1}{2-\alpha}}
\)
has units of viscosity (say Pa·s).

Figure~9 plots the normalized power spectra (PSD) of Brownian motion of 
particles immersed in a subdiffusive material with phenomenological 
material constant \(\mu_{\alpha}\) as a function of the dimensionless frequency 
\(\omega \lambda = \omega 
\left(\frac{m_{R}}{\mu_{\alpha}}\right)^{\!\tfrac{1}{2-\alpha}}\)
for different values of the fractional exponent \(\alpha\) of the power law 
given by equation~(52).

For the limiting case of \(\alpha = 1\), the springpot element becomes a 
Newtonian dashpot with \(\mu_{\alpha} = \mu_{1} = \eta\), and the power spectrum 
offered by equation~(62) reduces to the power spectrum given by 
equation~(15), since $\big[\mu_{\alpha} m_{R}^{\,1-\alpha}\big]^{\tfrac{1}{2-\alpha}}
   = 
\big[\eta\, m_{R}^{\,0}\big]^{\tfrac{1}{2-1}}
   = \eta.$

\begin{figure}[t]
    \centering
    \includegraphics[width=\columnwidth]{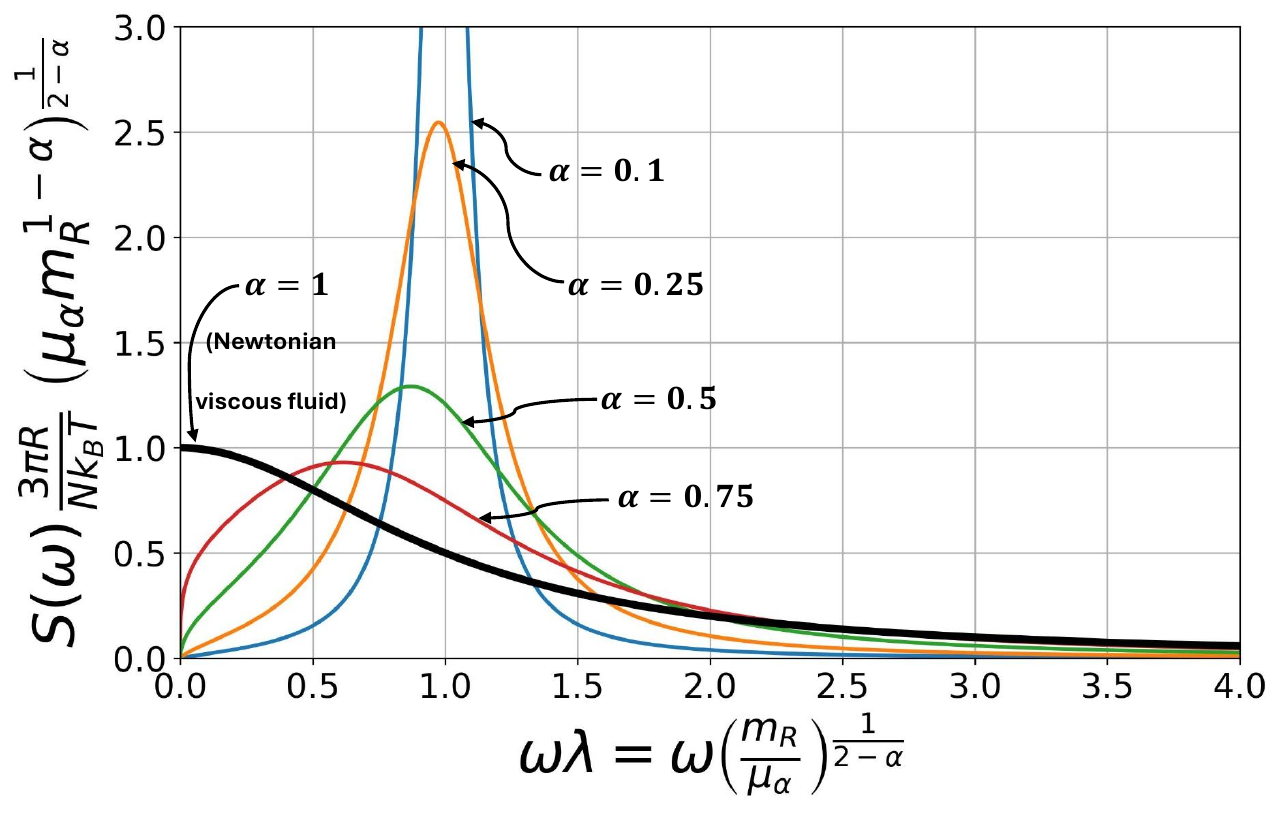}
    \caption{Normalized power spectra of Brownian motion within a subdiffusive material 
for different values of the fractional exponent $\alpha$, as a function of the 
dimensionless frequency 
$\omega\lambda = \omega\left(\frac{m_R}{\mu_\alpha}\right)^{\frac{1}{2-\alpha}}$.
}
\end{figure}

\section*{Brownian Motion in a Dense Viscous Fluid that Gives Rise to Hydrodynamic Memory}

When the density of the fluid surrounding a Brownian particle is appreciable, 
there is a convoluted interplay between the unsteady motion of the particle 
and the motion of the displaced fluid. Accordingly, in addition to the forces 
appearing in equation~(2), the Brownian particle also displaces the fluid in its 
immediate vicinity and in return, the fluid acts back on the particle and gives rise to long–range 
correlations that are different than the delta–correlations of Eq.~(4). 
The interaction of the Brownian particle with the displaced fluid can be 
described by the addition of inertia and memory terms to the basic Langevin 
equation~(2)~\cite{landau1980fluid,zwanzig1970,hinch1975,boussinesq1885},
\begin{equation}
\begin{aligned}
m \frac{dv(t)}{dt}
&= -\,\frac{2}{3}\pi R^{3}\rho_{f}\,\frac{dv(t)}{dt}
   - 6\pi R \eta\, v(t)\\
 &  - 6 R^{2}\sqrt{\pi \rho_{f}\eta}
     \int_{0}^{t} \frac{d v(\xi)/d\xi}{(t-\xi)^{1/2}}\, d\xi
   + f_{R}(t),
\end{aligned}
\tag{63}
\end{equation}
The quantity $\tfrac{2}{3}\pi R^{3}\rho_{f}$ in the additional inertia 
term of Eq.~{(63)} is half the mass of the displaced fluid 
$\tfrac{m_{f}}{2}$ ( $m_{f}=\tfrac{4}{3}\pi R^{3}\rho_{f}$ with 
$\rho_{f}$ = density of the surrounding fluid). whereas the following term 
$6\pi R \eta\, v(t)$ is the ordinary Stokes friction force, also present in 
Eq.~(2). The complete solution for the drag force on a sphere moving in a 
Newtonian viscous fluid in an arbitrary manner was first published in 
Boussinesq’s seminal paper~\cite{boussinesq1885}. The convolution integral term in 
Eq.~{(63)} is a hydrodynamic memory force that emerges from the 
reaction of the displaced fluid on the moving Brownian particle.

By recalling that the Riemann--Liouville fractional integral of a continuous 
function is defined by Eq.~(53), where the lower limit $c=0$, the hydrodynamic memory integral in Eq.~(63) is essentially the 
fractional integral of order $\tfrac{1}{2}$ of the acceleration history 
$\dfrac{dv(t)}{dt}$ of the Brownian particle. Accordingly, by virtue of 
Eq.~(53), Eq.~(63) is expressed as~[25]
\begin{equation}
\begin{aligned}
    M\,\frac{dv(t)}{dt}
    + 6\pi R\eta\,v(t)
   & + 6R^{2}\sqrt{\pi \rho_{f}\eta}\,
      \Gamma\!\left(1/2\right)
      I^{1/2}\!\left[\frac{dv(t)}{dt}\right]\\
   & = f_{R}(t),
\end{aligned}
    \tag{64}
\end{equation}
where $M = m + \tfrac{1}{2}m_{f}
    = \tfrac{4}{3}\pi R^{3}\left(\rho_{p} + \tfrac{1}{2}\rho_{f}\right),$ with $\rho_{p}$ the mass density of the Brownian particle.  
Using the properties of fractional calculus, the fractional integral of 
order $\tfrac{1}{2}$ of the derivative of the velocity (first derivative) 
is the fractional derivative of order $\tfrac{1}{2}$ applied to the 
velocity history:
\begin{equation}
    I^{1/2}\!\left[\frac{dv(t)}{dt}\right]
    = \frac{d^{1/2}v(t)}{dt^{1/2}}
    = \frac{1}{\Gamma\!\left(-\tfrac{1}{2}\right)}
      \int_{0}^{t}
      \frac{v(\xi)}{(t-\xi)^{3/2}}\,d\xi .
    \tag{65}
\end{equation}

More generally, the fractional derivative of order $\alpha \in \mathbb{R}^{+}$ 
of a continuous function $f(t)$ is defined, within the context of 
generalized functions, as the convolution of $f(t)$ with the kernel $\frac{1}{\Gamma(1-q)}\,\frac{1}{\Gamma^{q-1}}$~\cite{oldham1974,miller1993,mainardi2010,makris2021fractional}. By employing the result of Eq.~(65) in association with 
$\Gamma\!\left(\tfrac{1}{2}\right)=\sqrt{\pi}$, Eq.~(64) simplifies to
\begin{equation}
    M\,\frac{dv(t)}{dt}
    + 6\pi R^{2} \sqrt{\rho_{f}\eta}\,
      \frac{d^{1/2}v(t)}{dt^{1/2}}
    + 6\pi R\eta\,v(t)
    = f_{R}(t).
    \tag{66}
\end{equation}

Equation~(66) offers the remarkable result that 
fractional differentials do not only appear when modeling 
phenomenological power-law relaxation of a wide range of viscoelastic materials~\cite{nutting1921,gemant1936,gemant1938,scottblair1944,scottblair1947}, but also they emerge naturally 
from the solution of continuum–mechanics equations as they result from 
fundamental conservation laws~\cite{landau1980fluid,zwanzig1970,widom1971,boussinesq1885}. The Langevin equation (63) or (66), which 
accounts for the hydrodynamic memory, was solved analytically by 
Widom~\cite{widom1971} after solving an integral equation for the velocity autocorrelation 
function of the Brownian particle in association with the appropriate 
long–range correlation of the random process. Building on Widom’s 
solution~\cite{widom1971}, Hinch~\cite{hinch1975} derived the following result for the velocity 
autocorrelation function, also presented in Refs.~\cite{clercx1992,li2013}:
\begin{equation}
    \langle v(0)v(t) \rangle 
    = \frac{N k_{B} T}{M(b-a)}
    \left[
        b e^{b^{2} t} \operatorname{Erfc}(b\sqrt{t})
        - a e^{a^{2} t} \operatorname{Erfc}(a\sqrt{t})
    \right],
    \tag{67}
\end{equation}
with
\begin{equation}
    a = \frac{z + \sqrt{z^{2} - 4\zeta M}}{2M}, and
    \qquad  
   b = \frac{z - \sqrt{z^{2} - 4\zeta M}}{2M},
    \tag{68}
\end{equation}
In this paper we follow the notation of Clercx and Schram~\cite{clercx1992}, 
with $\zeta =6 \pi R \eta$ and $z = 6\pi R^{2} \sqrt{\rho_{f}\eta}$ with units 
\(
    [M][T]^{-1/2} 
\),
that is the coefficient of the \(1/2\) fractional derivative of the 
velocity of the Brownian particle appearing in the Langevin equation~(68).

The rheological analogue for Brownian motion with hydrodynamic memory 
was constructed by Makris~\cite{makris2021} by synthesizing the 
mechanical elements appearing on the left–hand side of the 
Langevin equation~(66). The rheological network consists of a dashpot 
with shear viscosity $\eta$, a fractional Scott–Blair element with 
material constant $\mu_{3/2} = R\sqrt{\rho_{f}\eta} = \frac{z}{6\pi R},$ and an inerter with distributed inertance $m_{R} = \frac{M}{6\pi R}
      = \frac{2}{9} R^{2}\rho_{f}\left(\frac{\rho_{p}}{\rho_{f}} + \frac{1}{2}\right),$ as shown in Fig.~10.

\par\noindent
Given the parallel connection of the dashpot, the inerpot, and the inerter,
the constitutive law of the rheological network shown in Fig. 10 is
\begin{equation}
\tau(t)
= \eta\,\frac{d\gamma(t)}{dt}
+ \mu_{3/2}\,\frac{d^{3/2}\gamma(t)}{dt^{3/2}}
+ m_R\,\frac{d^{2}\gamma(t)}{dt^{2}}.
\tag{69}
\end{equation}
The Fourier transform of Eq. (69) gives $\tau(\omega)
= G(\omega)\,\gamma(\omega)
= \left[ i\omega\eta
+ \mu_{3/2}(i\omega)^{3/2}
- m_R \omega^{2} \right]\gamma(\omega).$
The complex dynamic fluidity of the rheological model
described by Eq. (69) and schematically shown in Fig. 10 is

\begin{equation}
\varphi(\omega)
=\frac{i\omega}{G(\omega)}
=\frac{1}{m_R}
\frac{1}{{\frac{\eta}{m_R}
+\frac{\mu_{3/2}}{m_R}(i\omega)^{1/2}
+i\omega}} .
\tag{70}
\end{equation}

\vspace{1ex}
\begin{figure}[t]
    \centering
    \includegraphics[width=\columnwidth]{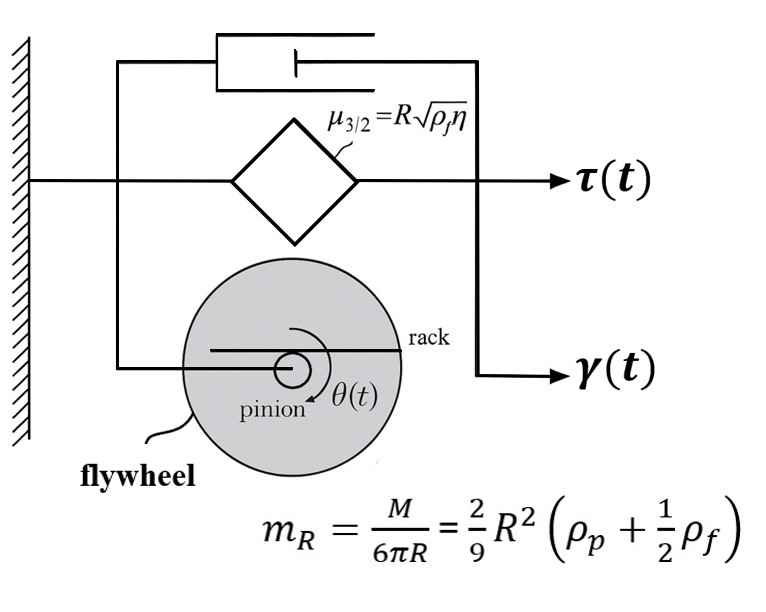}
    \caption{Rheological analogue for Brownian motion in a Newtonian viscous fluid with
  hydrodynamic memory. A dashpot with viscosity $\eta$ is connected in parallel with
  a Scott--Blair element with material constant $\mu_{3/2}=R\sqrt{\rho_f \eta}$ and an
  inerter with distributed inertance
  $m_R = \dfrac{M}{6\pi R}
       = \dfrac{2}{9} R^{2}\rho_f\!\left(\dfrac{\rho_p}{\rho_f}+\dfrac{1}{2}\right)$. (the dashpot–inerpot–inerter parallel connection)}
\end{figure}

where \((i\omega)^{1/2}=\omega^{1/2}\big(\cos \tfrac{\pi}{4}+i\sin \tfrac{\pi}{4}\big)\).  
Accordingly, the real part of Eq.~(70) is
\begin{widetext}
\begin{equation}
\Re\{\varphi(\omega)\}
=\frac{1}{m_R}
\frac{\dfrac{\eta}{m_R}
+\dfrac{\mu_{3/2}}{m_R}\,\omega^{1/2}\cos\tfrac{\pi}{4}}
{\left[\dfrac{\eta}{m_R}
+\dfrac{\mu_{3/2}}{m_R}\,\omega^{1/2}\cos\tfrac{\pi}{4}\right]^2
+
\left[\omega+\dfrac{\mu_{3/2}}{m_R}\,\omega^{1/2}\sin\tfrac{\pi}{4}\right]^2}
\tag{71}
\end{equation}
\end{widetext}

By recognizing that the quantity $\lambda=\left(\frac{m_R}{\mu_{3/2}}\right)^2$ has units of time, and upon multiplying the numerator and denominator
of Eq.~(71) with \(\lambda^2\) together with that $\sqrt{\lambda}\,\frac{\mu_{3/2}}{m_R}=1$, the general expression for the power spectral density (PSD)
given by Eq.~(24) becomes
\begin{widetext}
\begin{equation}
S(\omega)
=\frac{N k_B T}{3\pi R}
\Re\{\varphi(\omega)\}
=\frac{N k_B T}{3\pi R}\,
\dfrac{m_R}{\mu_{3/2}^{2}}
\frac{
\dfrac{\eta m_R}{\mu_{3/2}^{2}}+(\omega \lambda)^{1/2}
\cos\tfrac{\pi}{4}
}{
\left[
\dfrac{\eta m_R}{\mu_{3/2}^{2}}
+\big(\omega \lambda\big)^{1/2}\cos\tfrac{\pi}{4}
\right]^2
+
\left[
\omega \lambda
+\big(\omega \lambda\big)^{1/2}\sin\tfrac{\pi}{4}
\right]^2
}
\tag{72}
\end{equation}
\end{widetext}
where the quantity $\frac{\mu_{3/2}^{\,2}}{m_R}$ has units of viscosity (Pa·s). Therefore, the quantity $,\frac{\eta m_R}{\mu_{3/2}^{2}}$ is dimensionless.

Figure~11 plots the normalized power spectrum (PSD) of Brownian motion
in a dense viscous fluid that gives rise to hydrodynamic memory, as a 
function of the dimensionless frequency
$ \omega \lambda = \omega\left(\frac{m_R}{\mu_{3/2}^{\,}}\right)^2 $for different values of the dimensionless density ratio $ \gamma 
= \frac{\eta\, m_R}{\mu_{3/2}^{\,2}}
= \dfrac{1}{9}\left(1 + \dfrac{2\rho_p}{\rho_f}\right).$ 
As an example, when melamine rasin microspheres with 
$\rho_p = 1570\,\text{kg/m}^3$ 
are immersed in water 
$(\rho_f \approx 1000\,\text{kg/m}^3)$, 
$\frac{\rho_p}{\rho_f} = 1.57$ 
and 
$\gamma 
= \frac{\eta\, m_R}{\mu_{3/2}^{\,2}}= 0.46$;
whereas when they are immersed in acetone with 
$\rho_f = 784\,\text{kg/m}^3$, 
$\frac{\rho_p}{\rho_f} = 2.00$ 
and 
$\gamma=
\frac{\eta\, m_R}{\mu_{3/2}^{\,2}}= 0.55$.

\vspace{1ex}
\begin{figure}[t]
    \centering
    \includegraphics[width=\columnwidth]{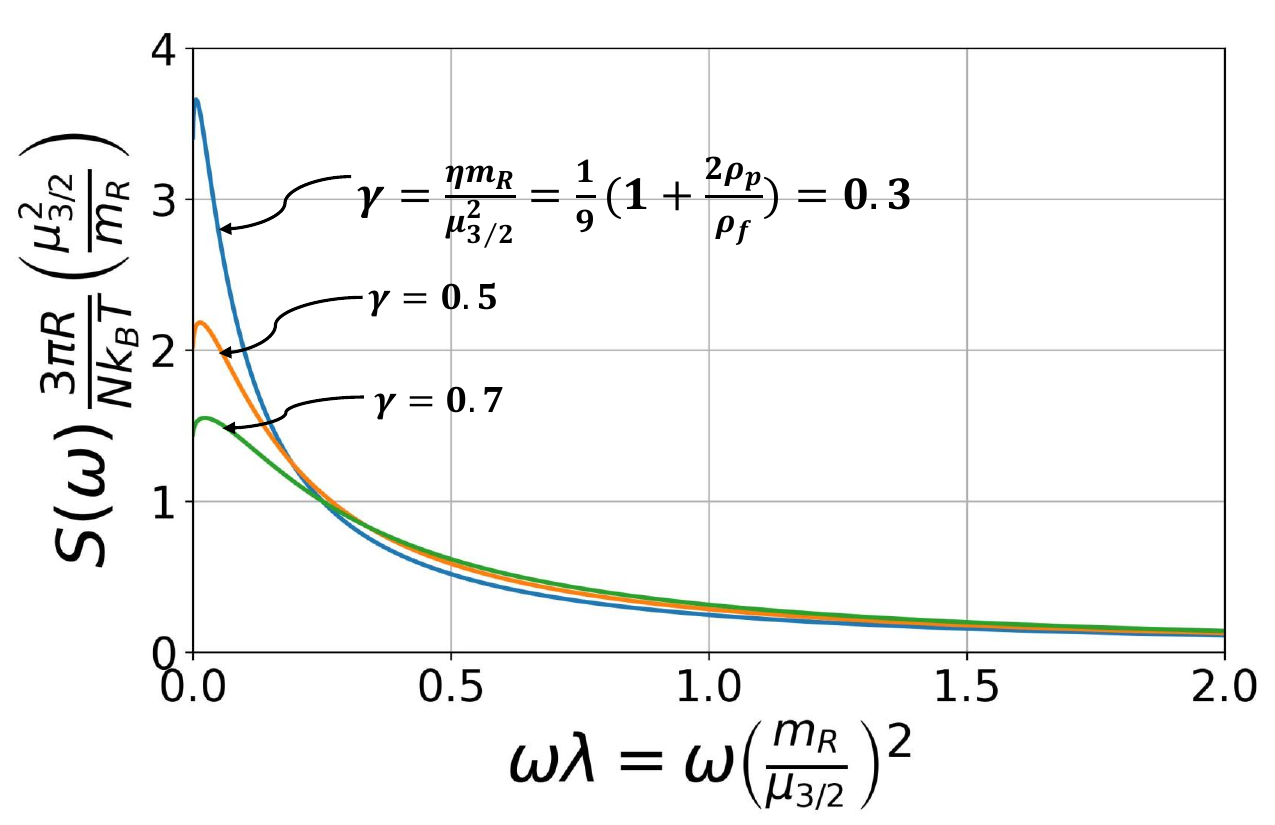}
    \caption{\textbf{Figure 11.} Normalized power spectra of Brownian motion within a 
dense viscous fluid that gives rise to hydrodynamic memory for 
different values of the dimensionless density ratio parameter $\gamma 
= \dfrac{\eta m_R}{\mu_{3/2}^{\,2}}
= \dfrac{1}{9}\left(1 + \frac{2 \rho_p}{\rho_f}\right),$ as a function of the dimensionless frequency $\omega \lambda = \omega\left(\frac{m_R}{\mu_{3/2}^{\,2}}\right)^2.$
}
\end{figure}

\newpage
\section*{Summary}

In this paper we employ a viscous--viscoelastic correspondence principle~\cite{makris2020,makris2021} for Brownian motion, and we show that the power spectrum (power spectral density) of the Brownian motion of microparticles of mass $m$ and radius $R$ immersed in any linear, isotropic viscoelastic material is proportional to the real part of the complex dynamic fluidity (complex mobility) of a linear rheological network that is a parallel connection of the linear viscoelastic material within which the Brownian particles are immersed and an inerter with distributed inertance $m_R = \frac{m}{6\pi R}$. The synthesis of this rheological analogue simplifies appreciably the calculations of the power spectrum of Brownian motion within viscoelastic materials.

Upon deriving the known results for the power spectrum of Brownian motion in a memoryless viscous fluid and within a linear Kelvin solid (harmonic trap), we present results for the power spectrum of Brownian motion within a Maxwell fluid, a Jeffreys fluid, a subdiffusive material and within a dense viscous fluid that gives rise to hydrodynamic memory.

\bibliographystyle{aipnum4-2}
\bibliography{refs}

\end{document}